\documentclass{aa}
\usepackage{graphicx}
\usepackage{txfonts}
\usepackage{natbib}
\usepackage{aalongtable,lscape}
\begin{document}
\titlerunning{The {\it XMM-Newton} Medium sensitivity Survey}
   \title{The {\it XMM-Newton} Serendipitous Survey}

   \subtitle{V. Optical identification of the {\it XMM-Newton} Medium sensitivity
   Survey (XMS)\thanks{Based on observations obtained with
    {\it XMM-Newton}, an ESA science mission with instruments and
    contributions directly funded by ESA Member States and the USA
    (NASA). Based on observations made with the INT/WHT, TNG and NOT operated on the island
    of La Palma by the Isaac Newton Group, the Centro Galileo Galilei and the Nordic Optical
    Telescope Science Association respectively, in the Spanish Observatorio del Roque de los Muchachos. Based on observations collected at the Centro Astron\'omico Hispano Alem\'an (CAHA) at Calar Alto,
    operated jointly by the Max-Planck Institut f\"ur Astronomie and the Instituto de Astrof\'\i sica de Andaluc\'\i a
    (CSIC). Based on observations collected at the European
    Southern Observatory, Paranal, Chile, as part of programme
    75.A-0336}
}

   \author{X. Barcons \inst{1}
          \and F.J. Carrera
          \inst{1}
          \and M.T. Ceballos
          \inst{1}
          \and M.J. Page
          \inst{2}
          \and J. Bussons-Gordo
          \inst{1}
          \and A. Corral
          \inst{1}
          \and J. Ebrero
          \inst{1}
          \and S. Mateos
          \inst{1,3}
          \and J.A. Tedds
          \inst{3}
          \and M.G. Watson
          \inst{3}
          \and D. Baskill
          \inst{3}
          \and M. Birkinshaw
          \inst{4}
          \and T. Boller
          \inst{5}
          \and N. Borisov
          \inst{6}
          \and M. Bremer
          \inst{4}
          \and G.E. Bromage
          \inst{7}
          \and H. Brunner
          \inst{5}
          \and A. Caccianiga
          \inst{8}
          \and C.S. Crawford
          \inst{9}
          \and M.S. Cropper
          \inst{2}
          \and R. Della Ceca
          \inst{8}
          \and P. Derry
          \inst{3}
          \and A.C. Fabian
          \inst{9}
          \and P. Guillout
          \inst{10}
          \and Y. Hashimoto
          \inst{5}
          \and G. Hasinger
          \inst{5}
          \and B.J.M. Hassall
          \inst{7}
          \and G. Lamer
          \inst{11}
          \and N.S. Loaring
          \inst{2,12}
          \and T. Maccacaro
          \inst{8}
          \and K.O. Mason
          \inst{2}
          \and R.G. McMahon
          \inst{9}
          \and L. Mirioni
          \inst{10}
          \and J.P.D. Mittaz
          \inst{2}
          \and C. Motch
          \inst{10}
          \and I. Negueruela
          \inst{10,13}
          \and J.P. Osborne
          \inst{3}
          \and F. Panessa
          \inst{1}
          \and I. P\'erez-Fournon
          \inst{14}
          \and J.P. Pye
          \inst{3}
          \and T.P. Roberts
          \inst{3,15}
          \and S. Rosen
          \inst{2,3}
          \and N. Schartel
          \inst{16}
          \and N. Schurch
          \inst{3,15}
          \and A. Schwope
          \inst{11}
          \and P. Severgnini
          \inst{8}
          \and R. Sharp
          \inst{9,17}
          \and G.C. Stewart
          \inst{3}
          \and G. Szokoly
          \inst{5}
          \and A. Ull\'an
          \inst{1,18}
          \and M.J. Ward
          \inst{3,15}
          \and R.S. Warwick
          \inst{3}
          \and P.J. Wheatley
          \inst{3,19}
          \and N.A. Webb
          \inst{20}
          \and D. Worrall
          \inst{4}
          \and W. Yuan
          \inst{9,21}
          \and H. Ziaeepour
          \inst{2}
          }

   \offprints{X. Barcons}

   \institute{Instituto de F\'\i sica de Cantabria (CSIC-UC), 39005
   Santander, Spain
         \and
         Mullard Space Science Laboratory, University College
         London, Holmbury St. Mary, Dorking, Surrey RH5 6NT, UK
         \and
         X-ray and Observational Astronomy Group, Department of Physics and Astronomy,
         Leicester University, Leicester LE1 7RH, UK
         \and
         H.H. Wills Physics Laboratory, University of Bristol, Tyndall Avenue, Bristol BS8
         1TL, UK
         \and
         Max-Planck-Institut f\"ur Extraterrestrische Physik,
         Giessenbachstrasse, 85740 Garching, Germany
         \and
         Special Astrophysical Observatory, 369167 Nizhnij Arkhyz, Russia
         \and
         Centre for Astrophysics, University of Central Lancashire, Preston PRI
         2HE, UK
         \and
         INAF - Osservatorio Astronomico di Brera, via Brera 28, 20121
         Milano, Italy
         \and
         Institute of Astronomy, Madingley Road, Cambridge CB3 0HA,
         UK
         \and
         Observatoire Astronomique de Strasbourg, 11 rue de
         l'Universit\'e, 67000 Strasbourg, France
         \and
         Astrophysikalisches Institut Potsdam, An der Sternwarte
         16, 144482, Potsdam, Germany
         \and
         South African Large Telescope, P.O. Box 9, Observatory,
         7935, South Africa
         \and
         Departamento de F\'\i sica, Ingenier\'\i a de Sistemas y Teor\'\i a de
         la Se\~nal, Universidad de Alicante, 03080 Alicante, Spain
         \and
         Instituto de Astrof\'\i sica de Canarias, 38200 La Laguna,
         Spain
         \and
         Department of Physics, University of Durham, South Road, Durham DH1
         3LE, UK
         \and
         European Space Astronomy Centre, Apartado 50727, 28080
         Madrid, Spain
         \and
         Anglo-Australian Observatory, PO Box 296, Epping, NSW 1710, Australia
         \and
         Centro de Astrobiolog\'\i a (CSIC-INTA), P.O. Box 50727,
         28080 Madrid, Spain
         \and
         Department of Physics, University of Warwick, Coventry CV4
         7AL, UK
         \and
         Centre d'Etude Spatiale des Rayonnements, CNRS/UPS,
         9 Avenue du Colonel Roche, 31028 Toulouse Cedex 04, France
         \and
         National Astronomical Observatories of China/Yunnan Observatory,
         Phoenix Hill, PO Box 110, Kunming, Yunnan, China
             }

   \date{Received 30 May 2007; accepted 2007}


  \abstract
  {}
   {X-ray sources at intermediate fluxes (a few
   $\times 10^{-14}\, {\rm erg\, cm^{-2}\, s^{-1}}$)
   with sky density of $\sim 100\, {\rm deg}^{-2}$, are responsible for
   a significant fraction of the cosmic X-ray
   background at various energies below 10 keV. The aim of this paper is to
   provide an unbiased and quantitative description of the
   X-ray source population at these fluxes and in various X-ray energy bands.}
   {We present the XMM-Newton Medium sensitivity Survey (XMS), including a
   total of 318 X-ray
   sources found among the serendipitous content of 25 XMM-Newton
   target fields. The XMS comprises four largely
   overlapping source samples selected at soft (0.5-2 keV),
   intermediate (0.5-4.5 keV), hard (2-10 keV) and ultra-hard (4.5-7.5 keV)
   bands, the first three of them being flux-limited.
   }
   {We report on the optical identification of the XMS samples, complete to
   85-95\%. At the flux levels sampled by the XMS
   we find that the X-ray sky is largely dominated by Active Galactic Nuclei.
   The fraction of stars in soft X-ray selected samples is below 10\%, and
   only a few per cent for hard
   selected samples. We find that the fraction of optically obscured objects
   in the AGN population stays constant at around 15-20\% for soft and
   intermediate band selected X-ray sources, over 2 decades of flux.
   The fraction of obscured objects amongst the AGN population is larger
   ($\sim 35-45\%$) in the hard or ultra-hard selected samples,
   and constant across a similarly wide flux range.
   The distribution in X-ray-to-optical flux ratio is a strong function of
   the selection band, with a larger fraction of sources with high values in
   hard selected samples.  Sources with X-ray-to-optical flux ratios in excess
   of 10 are dominated by obscured AGN, but with a significant
   contribution from unobscured AGN. }
   {}

   \keywords{X-rays:general, galaxies, stars; Galaxies: active}

   \maketitle
%

\section{Introduction}

Supermassive black holes (SMBHs, i.e., with masses $\sim 10^6-10^9\,
{\rm M}_{\odot}$) have been detected in the centers of virtually all
nearby galaxies \citep{Merrit01,Tremaine02}. In many of these
galaxies -including our own-, the SMBH is largely dormant, i.e., the
luminosity is many orders of magnitude below the Eddington limit.
Only $\sim 10\%$ of today's galaxies (at most) host active galactic
nuclei (AGN), and a very large fraction of them are in fact
inconspicuous at most wavelengths because of obscuration
\citep{Fabian99}.


It is generally believed that the seeds of these SMBHs were the
remnants of the first generation of massive stars in the history of
the Universe.  These early black holes may have had masses of tens
of ${\rm M}_{\odot}$ at most. The growth of these relic black holes
to their current sizes is very likely dominated by accretion, with
additional contributions by other phenomena like black hole mergers
and tidal capture of stars \citep{Marconi04}. According to current
synthesis models, the integrated
  X-ray emission produced by the growth of SMBHs by accretion over the
  history of the Universe is recorded in the X-ray background
  (XRB). Thus the XRB can be used to constrain the epochs and
  environments in which SMBHs developed.

There are currently a number of existing or on-going surveys in
various X-ray energy bands (see \citet{Brandt05} for a recent
compilation). In the pre-Chandra and pre-{\it XMM-Newton} era the
Einstein Extended Medium Sensitivity Survey
\citep{Maccacaro82,Gioia90,Stocke91} pioneered the procedure of
determining typical X-ray to optical flux ratios for different
classes of X-ray sources to facilitate the identification processes
and has set the standards for serendipitous X-ray surveys. $ROSAT$
produced a number of surveys in the soft 0.5-2 keV X-ray band at
various depths, e.g., the $ROSAT$ Bright Survey \citep{Schwope00},
the intermediate flux RIXOS survey \citep{Mason00} and the $ROSAT$
deep surveys
\citep{McHardy98,Georgantopoulos96,Hasinger98,Lehmann01} among
others. These surveys show that AGN dominate the high Galactic
latitude soft X-ray sky at virtually all relevant fluxes. The
majority of these AGN are of spectroscopic type 1, which means that
we are witnessing the growth of SMBH through unobscured lines of
sight. In a moderate fraction of the sources identified, however,
there is evidence for obscuration as their optical spectra lack
broad emission lines (type 2 AGN).

\citet{Ueda03} discuss the results from a large area X-ray survey in
the 2-10 keV band with ASCA and those from HEAO-1 and $Chandra$,
where a larger fraction of the sources identified correspond to type
2 AGN.

With {\it Chandra} and {\it XMM-Newton} coming into operation X-ray
surveys, particularly at energies above a few keV, have been
significantly boosted. Thanks to the high sensitivity and large
field of view of the EPIC cameras \citep{Turner01,Struder01} on
board {\it XMM-Newton} \citep{Jansen01}, X-ray surveys requiring
large solid angles have been dominated by this instrument.  The
Bright Source Survey-BSS \citep{Dellaceca04} contains 400 sources
brighter than $\sim 7\times 10^{-14}\, {\rm erg}\, {\rm cm}^{-2}\,
{\rm s}^{-1}$ either in 0.5-4.5 keV or 4.5-7.5 keV. The BSS
samples\footnote{{\tt http://www.brera.mi.astro.it/\~{}xmm/}}, which
have been identified to $\sim 90\%$ \citep{Caccianiga07}, show an
X-ray sky dominated by AGN, where the fraction of obscured objects
varies with the selection band (sample selection at harder energies
reveals a higher fraction of obscured objects as expected).

Deep surveys have also been conducted by {\it XMM-Newton}, for
example in the Lockman Hole down to $\sim 10^{-15} \, {\rm erg}\,
{\rm cm}^{-2}\, {\rm s}^{-1}$ \citep{Hasinger01,Mateos05b}. However,
thanks to its much better angular resolution, the {\it Chandra} deep
surveys are photon counting limited and far from confusion and are
consequently much more competitive at fainter fluxes
\citep{Alexander03,Tozzi06}. Optical identification of these deep
surveys is largely incomplete, a fact that is driven by the
intrinsic faintness and red colour of most of the counterparts to
the faintest X-ray sources. In the intermediate flux regime,
however, the identified fractions are large and nearing completion.
It is interesting to note that deep surveys start to find a
population of galaxies not necessarily hosting active nuclei as an
important ingredient. In addition, the AGN population is found to
contain an important fraction of obscured objects.

The wide range of intermediate X-ray fluxes, between say $10^{-15}\,
\, {\rm erg}\, {\rm cm}^{-2}\, {\rm s}^{-1}$ and $10^{-13}\, {\rm
erg}\, {\rm cm}^{-2}\, {\rm s}^{-1}$ have also been the subject of a
number of on-going surveys. Besides bridging the gap between wide
and deep surveys, intermediate fluxes sample the region around the
break in the X-ray source counts \citep{Carrera07}, and therefore
their sources are responsible for a large fraction of the X-ray
background. Among these, we highlight the {\it XMM-Newton} survey in
the well-studied (at many bands) COSMOS field, which covers $2\,
\deg^2$ to fluxes $\sim 10^{-15}\, {\rm erg\,  cm^{-2}\,  s^{-1}}$
\citep{Hasinger07}. The optical identification is still on-going,
reaching 40\% \citep{Brusa07}. At fluxes around $10^{-14}\, {\rm
erg}\, {\rm cm}^{-2}\, {\rm s}^{-1}$, the HELLAS2XMM survey
\citep{Baldi02,Fiore03}, now extended to cover $1.4\, \deg^2$,
contains over 220 X-ray sources, optically identified to 70\%
completeness \citep{Cocchia07}.

Other surveys in this flux range include the {\it XMM-Newton} survey
in the Marano field \citep{Krumpe07}, which is 65\% identified over
a modest solid angle of $0.28\, \deg^2$. Also the XMM-2dF survey
(Tedds et al., in preparation), which contains almost 1000 X-ray
sources optically identified in the Southern Hemisphere, is an
important contributor in this regime. {\it Chandra} has also
triggered surveys at intermediate fluxes, most notably the {\it
Chandra} Multiwavelength Survey \citep{Kim04a,Kim04b,Green04},
covering $1.7\, \deg^2$ and identified to $\sim 40\%$ completeness
\citep{Silverman05}.

In the realm of this variety of X-ray surveys that yield a
qualitative picture of the X-ray sky, the {\it XMM-Newton} Medium
sensitivity Survey (XMS) discussed in this paper, finds its role in
three important ways: a) it deals with very large samples, selected
at various X-ray bands where {\it XMM-Newton} is sensitive, from 0.5
to 10 keV; b) the samples that we consider have been identified
almost in full, from 85\% to 95\% completeness and c) three out of
the four samples that we explore are strictly flux limited in three
energy bands (0.5-2 keV, 0.5-4.5 keV and 2-10 keV). Armed with these
unique features, the XMS is a very powerful tool to derive a {\it
quantitative} characterization of the population of X-ray sources
selected in various bands, and also to study and characterize
minority populations, all of it at specific intermediate X-ray
fluxes where a substantial fraction of the X-ray background below 10
keV is generated. The power of the XMS is enhanced by the fact that
to some extent it is a representative sub-sample of the {\it
XMM-Newton} X-ray source catalogue 2XMM\footnote{Pre-release under
{\tt http://xmm.vilspa.esa.es/xsa}}, containing 150,000 entries.

Specific goals that have driven the construction of the XMS whose
results are presented in this paper include: a) quantify the
fraction of stars versus extragalactic sources at intermediate X-ray
fluxes and at different X-ray energy bands; b) quantify the fraction
of AGN that are classified as obscured by optical spectroscopy at
intermediate X-ray fluxes and for samples selected in different
energy bands; c) find the redshift distribution for the various
classes of extragalactic sources and compare soft and hard X-ray
selected samples; d) study the distribution of the X-ray-to-optical
flux ratio for the various classes of X-ray sources, also as a
function of X-ray selection band. The X-ray spectral properties of
the sources of the XMS were already discussed in \citet{Mateos05a}.

Further goals that we will achieve with the XMS in forthcoming
papers include: e) determine the fraction of ``red QSOs'' at
intermediate X-ray fluxes and as a function of X-ray selection band;
f) relate X-ray spectral properties (like photoelectric absorption)
to optical colours of the counterpart; g) quantify the fraction of
radio-loud AGN in the samples selected at various X-ray energies; h)
construct Spectral Energy Distributions for the various classes of
sources in the XMS.  Results on these further aspects will be
presented in a forthcoming paper (Bussons-Gordo et al., in
preparation).

The paper is organized as follows: in Section~\ref{sec:XMS} we
define the XMS along with the 4 samples that constitute it,
including the X-ray source list; in Section~\ref{sec:imaging} we
discuss the multi-band optical imaging conducted on the {\it
XMM-Newton} target fields and the process for selecting candidate
counterparts; this is continued in Section~\ref{sec:identification}
where we discuss the identification of the XMS sources in terms of
optical spectroscopy, and list photometric and spectroscopic
information on each XMS source. Section~\ref{sec:XMSpopulations}
presents the first scientific results from the XMS, specifically a
description of the overall source populations, the fraction of stars
in the various samples, the fraction of optically obscured AGN, and
the X-ray to optical flux ratio of the different source populations.
Section~\ref{sec:conclusions} summarizes our main results.

To clarify the terminology used in this paper, an AGN not displaying
broad emission lines in its optical spectrum is termed as type 2 or
obscured, and type 1 or unobscured otherwise.  The property of being
absorbed or unabsorbed refers only to the detection or not of
photoelectric X-ray absorption. Throughout this paper, we used a
single power law X-ray spectrum to convert from X-ray source count
rate to flux in physical units, with a photon spectral index
$\Gamma=1.8$ for the XMS-S and XMS-X samples and $\Gamma=1.7$ for
the XMS-H and XMS-U samples. These are the average values obtained
by \citet{Carrera07}, which -as opposed to what we do here- used the
specific value of $\Gamma$ for each individual source and energy
range. When computing luminosities, we also use the above spectra
for K-correction and the concordance cosmology parameter values:
$H_0=70\, {\rm km}\, {\rm s}^{-1}\, {\rm Mpc}^{-1}$, $\Omega_m=0.3$
and $\Omega_\Lambda=0.7$. All quoted uncertainties in parameter
estimates are shown at 90\% confidence level for one interesting
parameter.

\section{The {\it XMM-Newton} Medium sensitivity Survey (XMS)}
\label{sec:XMS}

The XMS is a serendipitous X-ray source survey with intermediate
X-ray fluxes, which has been built using the AXIS\footnote{AXIS (An
{\it XMM-Newton} International Survey) was an International Time
Programme of the Observatorio del Roque de Los Muchachos , which was
granted observing time in years 2000 and 2001. See {\tt
http://venus.ifca.unican.es/\~{}xray/AXIS} for details} sample
described in \citet{Carrera07}.  The XMS uses 25 target fields (see
table~\ref{WFCobs}, areas around targets themselves are excluded),
which cover a geometric sky area $\sim 3\, \deg^2$. The details of
the source searching, screening, masking out of problematic detector
areas (CCD gaps, bright targets, bad pixels and columns and out of
time events) are extensively discussed in \citet{Carrera07}.

\begin{table*}
\centering{
 \caption{XMS target fields}
\begin{tabular}{l c c c c}
\hline\hline
Target Field       &   RA (J2000)& DEC (J2000) & $b^{II}(\deg)$ & Phot$^a$\\
\hline
\object{Cl 0016+1609}     & 00:18:33 &+16:26:18 & -45.5 & SDSS\\
\object{G 133-69} pos2    & 01:04:00 &-06:42:00 & -69.3 & CMC\\
\object{G 133-69} pos1    & 01:04:24 &-06:24:00 & -68.7 & CMC\\
SDS-1b          & 02:18:00 &-05:00:00 & -59.7 & CMC\\
SDS-3           & 02:18:48 &-04:39:00 & -59.3 & CMC\\
SDS-2           & 02:19:36 &-05:00:00 & -58.9& CMC\\
\object{A 399}            & 02:58:25 &+13:18:00 & -39.2& CMC\\
\object{Mrk 3}            & 06:15:36 &+71:02:05 & +22.7 & WFC$^b$\\
\object{MS 0737.9+7441}   & 07:44:04 &+74:33:49 & +29.6 & WFC\\
S5 0836+71      & 08:41:24 &+70:53:41 & +34.4 & WFC\\
\object{Cl 0939+4713}     & 09:43:00 &+46:59:30 & +48.9 & SDSS\\
\object{B2 1028+31}       & 10:30:59 &+31:02:56 & +59.8 & SDSS\\
\object{B2 1128+31}       & 11:31:09 &+31:14:07 & +72.0 & SDSS\\
\object{Mrk 205}          & 12:21:44 &+75:18:37 & +41.7 &WFC\\
\object{MS 1229.2+6430}   & 12:31:32 &+64:14:21 & +53.0 & SDSS\\
\object{HD 117555}        & 13:30:47 &+24:13:58 & +80.7 & SDSS\\
\object{A 1837}           & 14:01:35 &-11:07:37 & +47.6 & CMC\\
\object{UZ Lib}           & 15:32:23 &-08:32:05 & +36.6 & WFC$^c$\\
\object{PKS 2126-15}      & 21:29:12 &-15:38:41 & -42.4 & CMC\\
\object{PKS 2135-14}      & 21:37:45 &-14:32:55 & -43.8 & CMC\\
\object{PB 5062}          & 22:05:10 &-01:55:18 & -43.3 & CMC\\
\object{LBQS 2212-1759}   & 22:15:32 &-17:44:05 & -52.9 & CMC\\
\object{PHL 5200}         & 22:28:30 &-05:18:55 & -50.0 & CMC\\
\object{IRAS 22491-1808}  & 22:51:50 &-17:52:23 & -61.4 & CMC\\
\object{EQ Peg}           & 23:31:50 &+19:56:17 & -39.1 & CMC\\
\hline\hline
\end{tabular}
}
\newline
\newline
\noindent $^a$ This column states what is the ultimate photometric
calibration used: WFC if our own data from a photometric night was
used as the ultimate resource, SDSS for the Sloan Digital Sky
Survey, and CMC for the Carlsberg Meridian Catalogue survey.

\noindent $^b$ In this case the extinction curve calibration in our
data gives some scatter which means that the magnitudes are not as
accurate as for the sources in the other fields.

\noindent $^c$ This field was not imaged in $g'$ and $r'$.  In
addition to $i'$, it was imaged in the Johnson filters B, V and R at
the ESO/MPG 2.2m telescope with the WFI camera.
 \label{WFCobs}
\end{table*}

The XMS itself is made of four largely overlapping samples.  The
XMS-S, XMS-H and XMS-X are flux limited in the 0.5-2 keV, 2-10 keV
and 0.5-4.5 keV bands respectively, with flux limits, well above the
sensitivity of the data, listed in Table \ref{completeness}. A
fourth sample (XMS-U) selected in the ``ultrahard'' band 4.5-7.5 keV
is not artificially limited in flux, and due to the scarcity of
these sources it contains all the sources detected in the 25 fields.
Table~\ref{Xraysources} lists the X-ray source positions and fluxes
in the various bands.
\addtocounter{table}{1}

The XMS-S and XMS-H were constructed to match the standard ``soft''
and ``hard'' X-ray bands that have been extensively used in previous
and contemporary X-ray surveys with {\it XMM-Newton} and other X-ray
observatories.  The 0.5-4.5 keV selection band of the XMS-X sample
was choosen to maximize the {\it XMM-Newton}/EPIC sensitivity and
indeed is the largest of the 4 samples.  The total number of
distinct X-ray sources in the XMS is 318, out of which 272 (86\%)
have been spectroscopically identified. The identification
completeness of the various samples is also shown in Table
\ref{completeness}, which exceeds 90\% for the softer XMS-S and
XMS-X samples, and is around 85\% for the hard XMS-H and ultra-hard
XMS-U samples. The XMS-X sample extends by an order of magnitude the
size of the pilot study presented in \citet{Barcons02}.

\section{Imaging and selection of candidate counterparts}
\label{sec:imaging}

\subsection{The data}

Target fields were observed primarily with the Wide-Field Camera
(WFC) on the 2.5m INT telescope. The observations were obtained via
the AXIS programme and other programmes devoted to image a large
number of {\it XMM-Newton} target fields in the optical. The WFC
covers virtually all the field of view of EPIC, if centered
optimally. We used the Sloan Digital Sky Survey filters $g'$, $r'$
and $i'$ to image all the XMS target fields.  In addition many of
the fields were also imaged bluewards and redwards using existing
facility filters at the WFC  ($u$ and $Z$ Gunn). These data are
available for all fields, except for the \object{G 133-69 pos 1} and
\object{PB 5062} fields, while for \object{UZ Lib} and \object{B2
1128+31} the $u$-band data are missing. Since data from these two
additional filters are not used in this paper, we do not discuss
them any further.

\begin{table}
\caption{Summary of identifications in the various samples}
\begin{tabular}{r c c c c l}
\hline\hline
Sample & Sel. band       & Limiting   & Number  & Number  & Fract. \\
       & (keV)          &   flux$^a$        & sources  & ident.              & (\%) \\ \hline
 XMS-S & 0.5-2.0         &   1.5        &  210         & 200            & 95    \\
 XMS-X & 0.5-4.5         &   2.0        &  284         & 261            & 92    \\
 XMS-H & 2.0-10        &   3.3        &  159         & 132            & 83   \\
 XMS-U & 4.5-7.5         &  1.15       &   70         &  60            & 86    \\
 \hline\hline
\end{tabular}
\newline

\noindent $^a$ in units of $10^{-14}\, {\rm erg}\, {\rm cm}^{-2}\,
{\rm s}^{-1}$ in the selection X-ray band\label{completeness}
\end{table}

Exposure times were adjusted to be deep enough for most of the X-ray
sources to have an optical counterpart in the $r'$ and $i'$ filters,
and therefore had to be significantly deeper than those in the
Digitized Sky Surveys. They were chosen as 600s, 600s and 1200s
respectively in the $g'$, $r'$, $i'$  filters for dark time.  This
produced images with limiting magnitude for point sources going down
to $r'\sim 23-24$ for $\sim 1-1.5"$ seeing, most typical in our
observing runs, which experience with the first fields
\citep{Barcons02} demonstrated to be appropriate.  When observing
with brighter moon conditions, we restricted ourselves to the
reddest filters and doubled the exposure times.

The WFC images were reduced using standard techniques including
de-bias, non-linearity correction, flat fielding and fringe
correction (in $i'$ and $Z$). Bias frames and twilight flats
obtained during the same observing nights were used, but for the
fringe correction, contemporaneous archival $i'$ and $Z$ fringe
frames were utilised. Information on the WFC pipeline procedures,
which performs all these steps can be found under the Cambridge
Astronomy Survey Unit\footnote{\tt
http://www.ast.cam.ac.uk/\~{}wfcsur} (CASU) web pages.

\subsection{Photometric calibration}

The photometric calibration of the WFC images was conducted in the
standard way. Photometric standard stars were observed during the
same nights as the {\it XMM-Newton} target fields were imaged, at
different air masses.  Then an extinction curve was fitted for each
optical band. In several cases were we suspected that photometric
conditions were not fulfilled by the original data, we re-imaged the
same field with one WFC filter ($r'$) or alternatively a part of it
with the ALFOSC instrument in imaging mode on the NOT telescope.

However, in a number of target fields and for some of the bands, the
extinction curve showed significant scatter that was attributed to
these observations being done under non-photometric conditions. In
order to improve the photometric quality of the data, two steps were
taken.  First we concentrated on calibrating one band (typically
$r'$) and later we applied colour corrections to propagate the
improved photometry to all bands.

\begin{figure}
   \centering
   \includegraphics[height=9cm,angle=270]{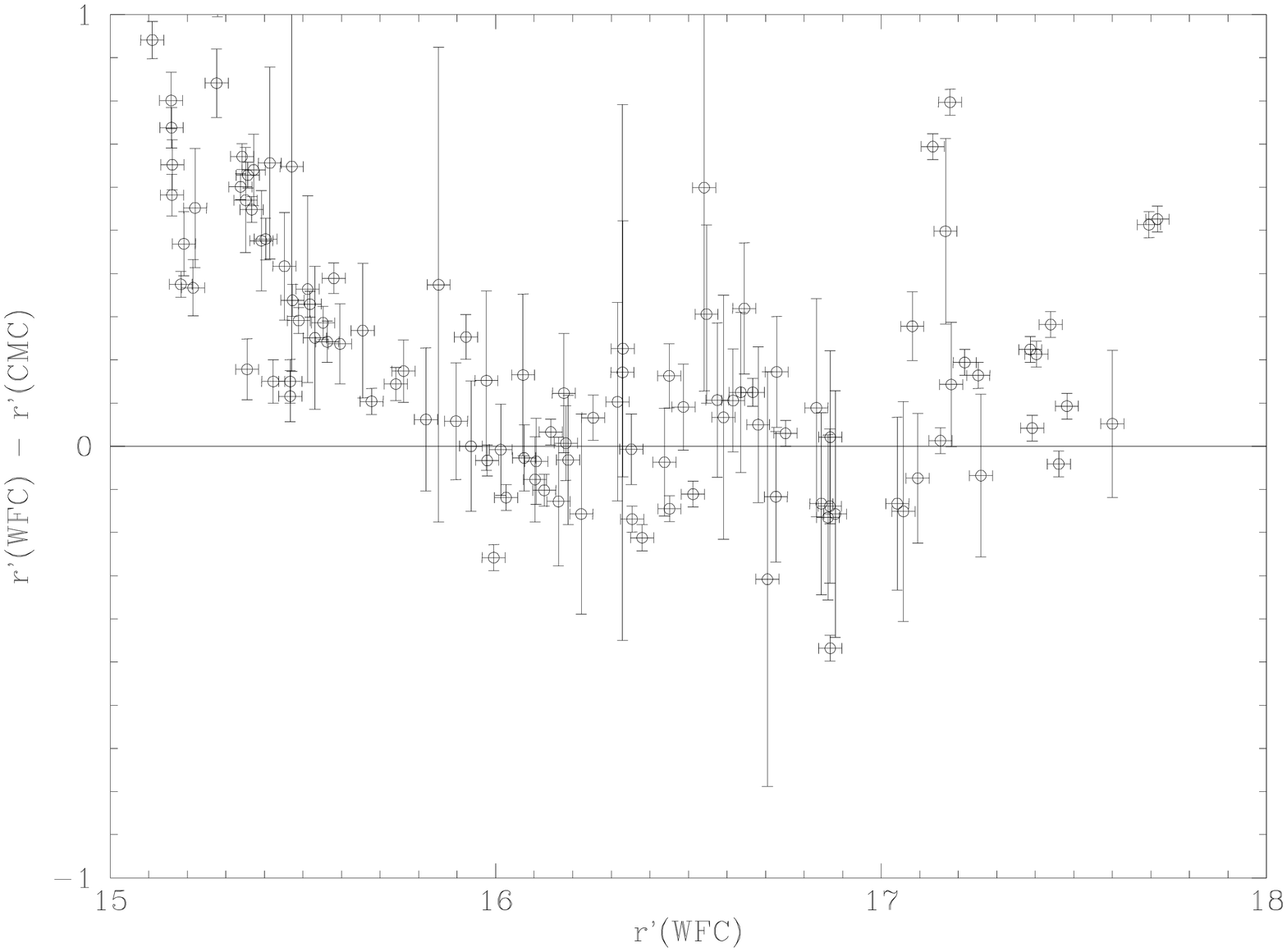}
   \includegraphics[height=9cm,angle=270]{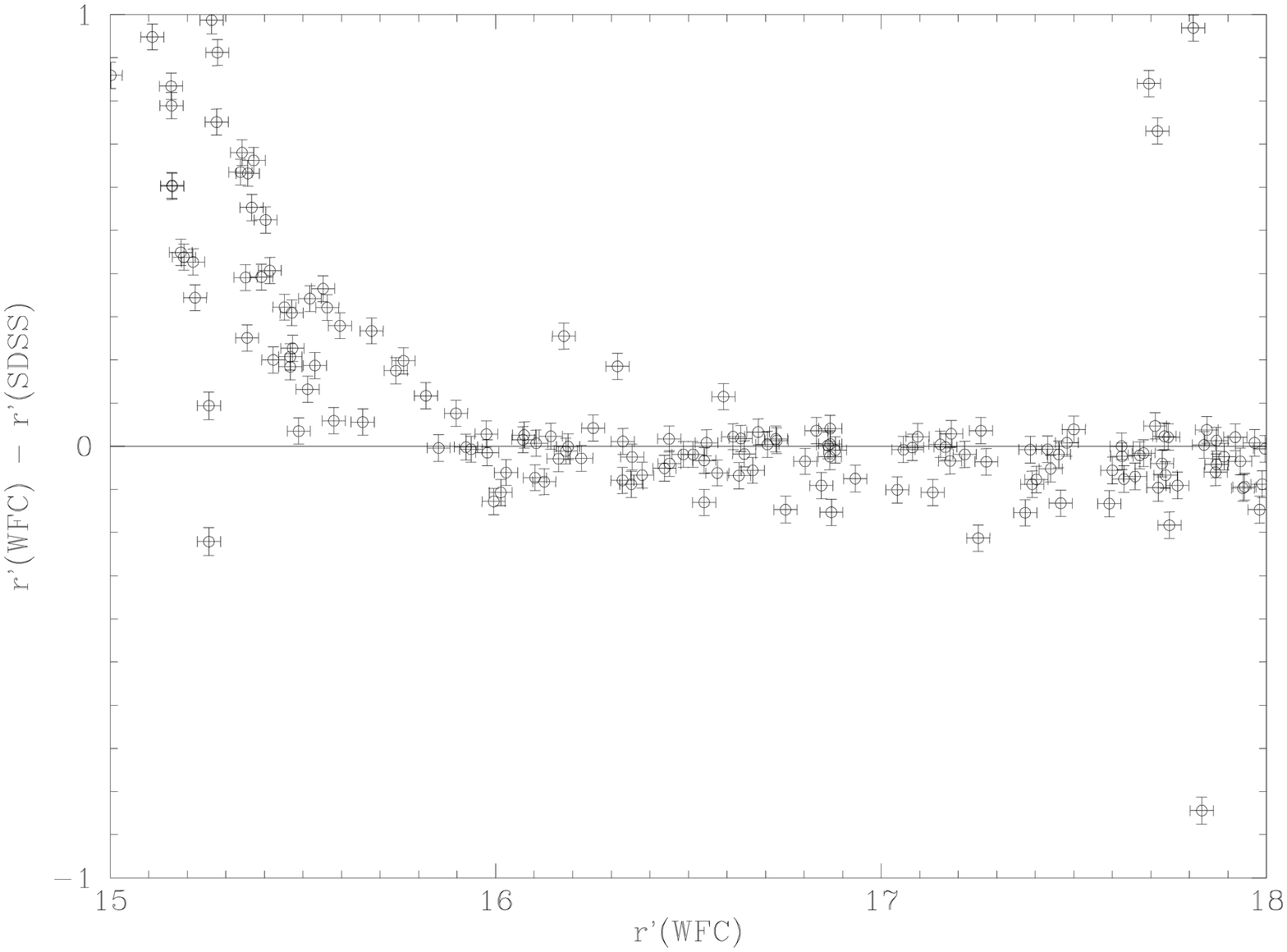}
      \caption{Photometric cross-calibration in the B2 1128+31 field, where we have both
      coverage from the SDSS and CMC, along with our own WFC photometry.}
         \label{photometry}
   \end{figure}

In the first step we used two complementary data sets
photometrically calibrated. The first of these is the Sloan Digital
Sky Survey, Data Release 5\footnote{{\tt http://www.sdss.org}}. We
could use the SDSS data on 6 {\it XMM-Newton} fields. The sky
density of SDSS is indeed lower than our WFC images, but still we
found typically a large enough number ($\sim 100$) of matches in
every image.

The second data set used to improve the photometric calibration is
the Carlsberg Meridian Catalogue (CMC) astrometric survey in the
$r'$ band\footnote{see {\tt
http://www.ast.cam.ac.uk/\~{}dwe/SRF/camc.html} for details}, which
we could apply to 19 fields. This survey is indeed much shallower
than the SDSS ($r'<17$). In addition we found very significant
systematic differences between WFC magnitudes and CMC ones at
magnitudes smaller than $r'\sim 16-16.5$ which we attributed to
saturation in our data.  That typically leaves a very narrow dynamic
range for cross-calibrating WFC versus CMC magnitudes, that we
adopted as $16.5<r'<17$.  Prompted by this, we also restricted the
WFC versus SDSS cross calibration to magnitudes brighter than
$r'\sim 16$. As a safety test, we cross-calibrated CMC versus
 SDSS $r'$ magnitudes in the 5 fields where we could do that, but in
 this case using the full magnitude range from $r'\sim 14-18$ and
 found tiny significant shifts, all of them well below $0.1\, {\rm mag}$ in
 all fields. Fig.~\ref{photometry} illustrates the residuals of the
cross-calibration in
 the case of one target field where we had all three WFC, SDSS and
 CMC data.

In general, photometric shifts in fields where the quality of the
WFC photometric calibration was thought to be good, were found to be
small (always $\Delta r'<0.2\, {\rm mag}$) when calibrated against
CMC or SDSS. In other cases where we had reasons to suspect that our
initial photometric calibration was not of high quality, we did find
indeed photometric shifts as large as $\Delta r'\sim 0.5\, {\rm
mag}$. This is why we decided to apply these corrections to our
photometry, with the SDSS one taking priority over CMC.
Table~\ref{WFCobs} lists the ultimate photometric calibration data
used in each field.

Armed with this refined calibration in $r'$, we then exported this
into the $g'$ and $i'$ bands by constructing a $g'-r'$ vs $r'-i'$
colour-colour diagram. We then compared this to a calibrated
colour-colour sequence pattern that was constructed using WFC
observations of ELAIS fields. Shifts were applied to $g'$ and $i'$
WFC magnitudes as to match both. These shifts were propagated to all
magnitudes listed in this paper.

The bottom line is that we believe our photometry to be better than
$0.1\, {\rm mag}$ in the majority of our fields and certainly better
than $0.2\, {\rm mag}$ for all of them.  In section 5.5, where we
analyze the X-ray-to-optical flux ratio, we use the quantity $\log
f_{opt}$, which has a maximum error due to these calibration
uncertainties well below 10\%.

\subsection{Astrometric calibration}

Astrometric calibration of the WFC was performed using the Cambridge
Astronomy Survey Unit (CASU) procedures. Typically, hundreds of
matches per WFC image were obtained against the APM
catalogue\footnote{\tt http://www.ast.cam.ac.uk/\~{}mike/apmcat/},
which was used as the astrometric reference for the optical images.
Specifically, a simple 6 parameter plate solution over the whole
4-CCD image was used, but accounting for a known and previously
calibrated telescope distortion cubic radial term. The residuals
from the plate solution, were typically below 0.2 arcsec, which is
indeed good enough for identifying candidate counterparts to the
X-ray sources and for blind spectroscopic observations to identify
these counterparts.

It should be noted that the astrometry of the {\it XMM-Newton} X-ray
source position was registered against the USNO A2 \citep{Monet98}
source catalogue \citep{Carrera07}, and that the optical astrometry
refers to a different astrometric system.  In order to ensure that
this does not lead to artificial mismatches, we measured the
USNO-APM shifts in every {\it XMM-Newton} field by cross-correlating
both source catalogues in the corresponding region.  The shifts were
significant in most cases but small, typically $< 0.5$~arcsec, which
is less than the statistical accuracy in the X-ray source positions
($0.6$~arcsec averaged over the whole XMS sample). The positions of
bright sources which were severely saturated in our WFC images were
obtained from the USNO catalogue itself and therefore do not suffer
from these small APM-USNO shifts. Given the small size of these
shifts and in view of the much broader overall distribution of
offsets between the position of the X-ray source and its optical
counterpart (see Fig.~\ref{offsets}) we conclude that the use of
these two different astrometric reference frames does not affect in
any noticeable way the results presented in this paper.

\section{Identification of the XMS sources}
\label{sec:identification}

In order to search for candidate counterparts of the X-ray sources,
we normally used the $r'$-band WFC image.  Optical source lists for
these images were generated with the CASU procedures.

Counterparts for the X-ray sources were searched for in the optical
image lists. Candidate counterparts had to be either within the 5
statistical errors (at 90\% confidence) of the X-ray position or
within 5 arcsec from the position of the X-ray source. This last
criterion was used to accommodate any residual systematics in the
astrometric calibration of the X-ray EPIC images.

\begin{figure}
   \centering
   \includegraphics[height=9cm,angle=270]{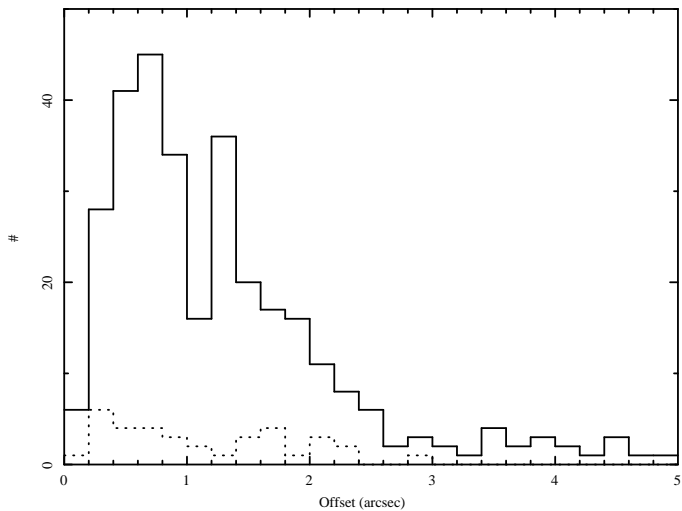}
      \caption{Histogram of the distances from the optical source to the X-ray source
      centroid.
      The continuous line is for all sources with a counterpart and
      the dotted line for those with a likely counterpart without
      spectroscopic confirmation.}
         \label{offsets}
   \end{figure}

As reported in \citet{Barcons02} this resulted in a vast majority of
the XMS sources having a single candidate counterpart. There are a
few exceptions to this.  In a few cases (15), the position of the
X-ray source happened to fall in the gaps between CCDs in the WFC
images. The strategy adopted to image in all optical filters with
the same aim point, which allowed us to obtain reliable optical
colour information for the vast majority of the sources, also
implied that for these few sources there is no optical image in any
of the optical filters covering the region around these X-ray
sources. In a fraction of these sources (14 out of 15), the
candidate counterpart was found by considering other optical imaging
data, mostly the USNO A2 catalogue, or complementary imaging data.

Also, in a modest amount of sources (78), there was more than one
single candidate counterpart formally complying with our proximity
criteria. But in 70 out of these 78 the optical source closest to
the X-ray source position was also brightest and we adopted that as
the likely counterpart. Given the brightness of the optical
counterparts $r'<22$ and the small region searched for around every
X-ray source, we are confident that the number of spurious
associations is insignificant in this sample.

Figure~\ref{offsets} shows the histogram of the X-ray to optical
angular separations for the sources spectroscopically identified and
for those where a unique candidate counterpart is found but without
a spectroscopic identification.  The distribution peaks at small
separations ($\sim 2"$).  Integrating this distribution outwards
shows that in 68\%, 90\% and 95\% of the cases the optical
counterpart lies closer than 1.5", 2.4" and 3.6" from the X-ray
source respectively. Although the histogram of unidentified sources
looks slightly more disperse than that of the identified ones, all
candidate optical counterparts fall within 3" from the position of
the X-ray source.

There are a total of five sources that have no candidate counterpart
in any of our optical images. For a further 3 sources, finding a
candidate counterpart required special strategy: in one case a
counterpart was only found in the $K$-band (XMSJ 122143.6+752238),
and in a further two cases the very faint optical counterparts were
only detected via imaging with the VLT (XMSJ 225227.6-180223, in the
I band) and Subaru (XMSJ 021705,4-045654, with $R=25.60$)
telescopes.

\subsection{Optical spectroscopy}

Searches for information on the XMS
  candidate counterparts in existing catalogues gave useful
  information (i.e., nature of the source and redshift) only for a
  handful of objects. Identifications
for a few other X-ray sources were provided to us by the Subaru/{\it
XMM-Newton} Deep Survey project (M. Akiyama, private communication)
and by the {\it XMM-Newton} Bright Source Survey
\citep{Dellaceca04}. That means that the vast majority of the XMS
sources were previously unidentified and required optical
spectroscopy.

Optical spectroscopy was conducted in a number of ground-based
optical facilities, following the strategy presented in
\citet{Barcons02}. The low surface density in the sky of the XMS
sources ($\sim 100\, {\rm deg}^{-2}$) makes the use of multi-object
slit-mask spectrographs not particularly efficient.  Part of the
identifications were performed using a fibre spectrometer
(AUTOFIB2/WYFFOS) which covers a much larger solid angle in the sky
and therefore is better suited for the identification work.

The main limitation of the fibre spectrometers in obtaining the
spectrum of faint sources resides in the subtraction of the sky
which enters the fibres along with the light from the target
objects. Wider fibres make this problem worst. This limits the
ultimate sensitivity of the spectrometer, which for our exposure
times and observing conditions was rarely good enough for magnitudes
fainter than $r'\sim 20.5$.

Therefore, despite the larger solid angle covered by fibre
spectrometers, the distribution of optical magnitudes in the XMS
source candidate counterparts calls for the use of single object
long-slit spectroscopy.  A number of such spectrometers were used in
a variety of ground-based telescopes, with apertures from 2.5m to
8.2m.

Table~\ref{SPECobs} lists the telescopes and observatories that were
used, along with the specific spectrometers, with specification of
the wavelength range, the slit width (or fibre width when
applicable) along with the measured spectral resolution using
unblended arc lines (or a sky line in the case of the fibre
spectrometer). The spectral reduction process is standard and was
described in \citet{Barcons02}. The final spectra will be available
under {\tt http://www.ifca.unican.es/\~{} xray/AXIS} and in the long
term in the {\it XMM-Newton} Science Archive\footnote{{\tt
http://xmm.vilspa.esa.es/xsa}} under the 2XMM catalogue.

It is worth recalling that these spectra are meant only to be
reliable for {\it identification} purposes, i.e., the
spectrophotometric calibration has only been performed at best in
relative terms (i.e., up to an absolute normalisation factor). Even
more, in the fibre spectra and in some of the long-slit spectra
which were not taken with the slit aligned to the parallactic angle,
differential refraction will cause the overall large-scale shape of
the spectrum to be incorrect. None of these facts hamper the
identification of the spectral features that we used in this paper,
which is based on broad and/or narrow emission lines and on
absorption bands, but not on broad-band features like the 4000~\AA\
break. However we want to caution against the use of these spectra
to measure line fluxes or line ratios because of the above
limitations.

\begin{table*}
\centering
 \caption{List of spectroscopic setups relevant to this
sample.}
\begin{tabular}{r c c c c l}
\hline
Telescope & Instrument       & Spectral       & Slit width  & Spectral              & Comments \\
          &                  & range (\AA )   & (arcsec)    & resolution$^a$ (\AA ) & \\ \hline
 WHT/ORM      & AUTOFIB2/WYFFOS  & 3900-7100      &  2.7$^b$    & 7                   & Fibre  \\
 WHT/ORM      & AUTOFIB2/WYFFOS  & 3900-7100      &  1.6$^b$    & 6                      & Fibre \\
 WHT/ORM      & ISIS             & 3500-8500      &  1.2-2.0    & 3.0-3.3               & Long slit\\
 TNG/ORM      & DOLORES          & 3500-8000      &  1.0-1.5    & 14-15                 & Long slit\\
 NOT/ORM      & ALFOSC           & 4000-9000      &  1.0-1.5    & 4                     & Long slit\\
 3.5m/CAHA    & MOSCA            & 3300-10000     &  1.0-1.7    & 24                      & Long slit\\
 UT1/ESO      & FORS2            & 4400-10000     &  1.0        & 6-12                      & Long slit\\

 \hline
\end{tabular}

$^a$ Measured from unsaturated arc lines\\
$^b$ Width of individual fibres \label{SPECobs}
\end{table*}

\subsection{Classification of the sources}

Based on the optical spectroscopy, we classify the counterparts to
the XMS X-ray sources as in \citet{Barcons02}.  Extragalactic
sources exhibiting broad emission lines (velocity widths in excess
of $\sim 1500\, {\rm km}\, {\rm s}^{-1}$) are classified as BLAGN
(Broad Line Active Galactic Nuclei); those exhibiting only narrow
emission lines are termed NELG (Narrow Emission Line Galaxies);
those with galaxy spectra without obvious emission lines are
classified in principle as Absorption Line Galaxies (ALG). Out of
the latter, we distinguish two classes of exceptions: two of the
sources with a galaxy spectrum without emission lines were
previously catalogued as BL Lac objects and we classify them as
such; if a qualitative inspection of the optical images show obvious
evidence for a galaxy concentration we then classify the source as a
cluster (Clus). Finally all X-ray sources with a stellar spectrum
are labeled simply as "Star".

This classification is simple to perform, but it lacks in some cases
a more detailed physical description of the source.  This is
particularly true in the case of the NELG, because no line
diagnostics are performed to check whether the object hosts an AGN
or not.  The reason is that due to the rather wide redshift range
spanned by these sources and the rather narrow wavelength coverage
of the optical spectra (particularly for the fibre spectroscopy) the
number of lines detected is small. Therefore it happens that typical
diagnostic lines drift out of the spectrum with redshift. In
addition, the quality of the spectra are in most cases not good
enough to detect weak lines necessary for these diagnostics.  In
fact the NELG are likely to be a mixed bag of type-2 AGN and star
forming galaxies.  In the discussion of the various samples we use
the X-ray luminosity as an indicator of the presence of an AGN in
these objects.

The second limitation of this simple classification is in the case
of clusters.  The very small number of X-ray sources identified as
clusters (2) is not a real property of the X-ray sky at these flux
levels, but an artifact of the detection and classification method.
The detection method for X-ray sources is designed for point sources
and might be missing a number of extended X-ray sources. In
addition, in several optical images, the presence of a cluster of
galaxies might not be obvious and the source might have been
identified as a galaxy with or without emission lines.

The third and final limitation is in the stellar content.  The stars
that we identify have a varied range of spectral properties, but
this is not explored in the present paper.

Redshifts have been measured by matching the most prominent features
in emission or absorption to sliding wavelengths of these features.
Templates for QSO and galaxies with a range of spectroscopic classes
were used to assist in the generation of first guesses when
necessary, especially when there were no prominent features.

Table \ref{IDtable} displays the full list of the 318 XMS sources,
along with optical magnitudes of their counterparts, and the
identifications.
\addtocounter{table}{1}

\section{The XMS X-ray source populations}
\label{sec:XMSpopulations}

\subsection{General}

\begin{figure*}
   \centering
   \includegraphics[height=16cm,angle=270]{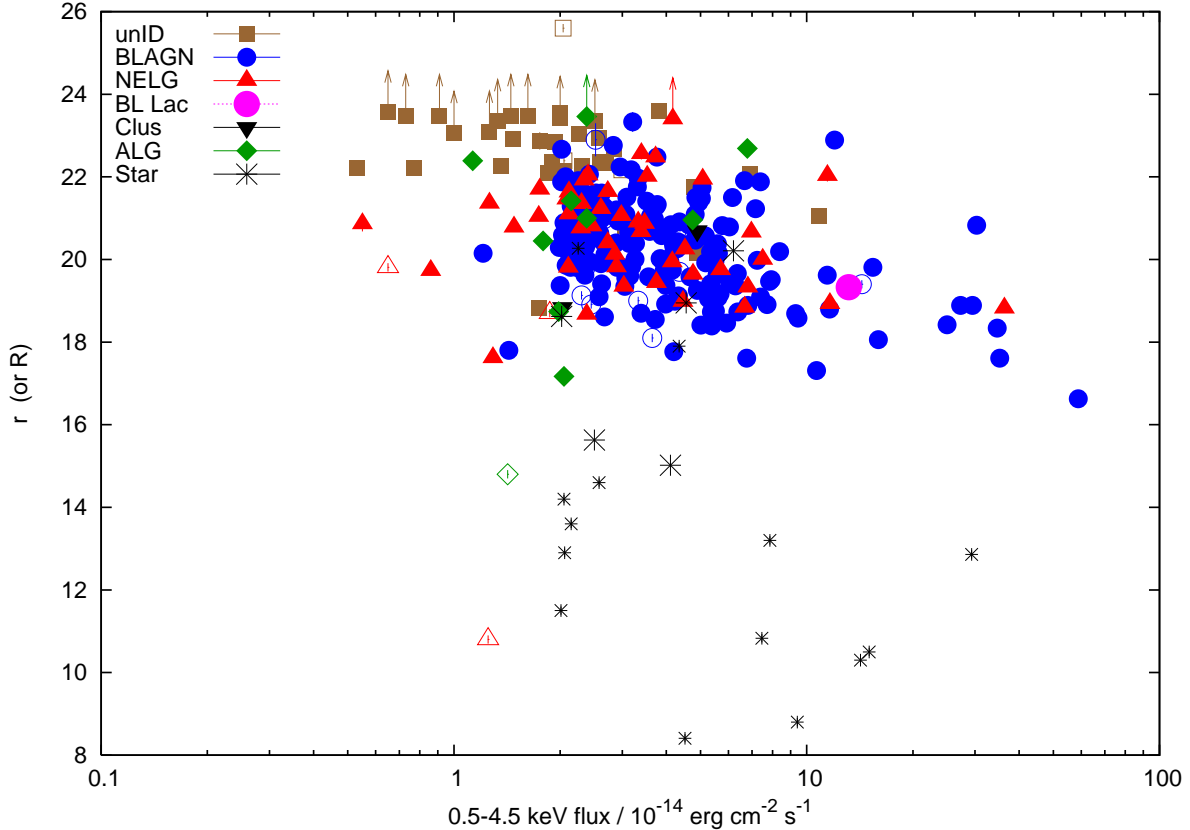}
      \caption{Optical magnitude versus 0.5-4.5 keV flux for the XMS sources.
      The optical magnitude shown is $r'$ (filled symbols) when available, and otherwise $R$ (hollow symbols).
      Upward arrows denote lower limits in the magnitude derived from the lack of optical counterparts in the WFC $r'$ band image, but
      note that several of these sources have counterparts in other optical bands.}
         \label{FXFopt_flux_all}
   \end{figure*}

The breakdown of the identifications in the 4 XMS samples is shown
in Table~\ref{idsummary}. The completeness of these identifications
is higher for the XMS-S (95\%) and XMS-X (92\%) than for the XMS-H
(83\%) and XMS-U (86\%). There are several reasons for that, the
most important one being that the XMS was originally conceived
around the 0.5-4.5 keV band to optimise the {\it XMM-Newton} EPIC
sensitivity and therefore the identification strategy has been
specially successful in this band.

In particular, and as we will show later, the {\it Hard} and {\it
Ultra-hard} samples contain a higher fraction of sources with a
higher X-ray-to-optical flux ratio and therefore more sources have
optically fainter counterparts. Given the limitations of the access
to 8-10m aperture class telescopes, in practice this means that the
identification incompleteness is also biased. This implies that the
fraction of unidentified sources is likely to be richer in
potentially obscured objects than the average.

The results for the XMS-X are particularly robust and in fact their
robustness can be verified by using what we might call the
"Southern" subset of the XMS-X.  This is due to the fact that
virtually all sources in this sample which are accessible from the
VLT at ESO were observed in September 2005 during the 075.A-0336 run
and the vast majority of them were identified. Table~\ref{idsummary}
also displays the numbers of identified targets in fields below a
declination of $+20^{\circ}$ in the XMS-X sample. In this sample, at
the price of reducing the size from the parent sample to $\sim
60\%$, we raise the identification fraction to over 98\% (only 4
sources out of 167 remain unidentified).

\begin{table*}
\centering
 \caption{\label{idsummary} Summary of the identifications
of the various XMS samples}
\begin{tabular}{l c c c c c c c c}
\hline\hline
Sample & Total & BLAGN & NELG & ALG & BL Lac & Clus & Star & Unid\\
\hline
XMS-S & 210 & 150 & 26 & 6 & 2 & 1 & 15 & 10\\
XMS-X & 284 & 192 & 38 & 7 & 2 & 2 & 20 & 23 \\
XMS-X (South) & 167 & 120 & 25 & 4 & 0 & 1 & 13 & 4\\
XMS-H & 159 & 85  & 34 & 7 & 2 & 1 & 3 & 27\\
XMS-U & 70 & 41 & 15 & 2 & 2 & 0 & 0 & 10\\
\hline
\end{tabular}
\end{table*}

A first glance at the overall source population that we are
sampling, can be seen in Fig.~\ref{FXFopt_flux_all}, where we have
plotted the optical magnitude (typically $r'$, but $R$ when $r'$ is
not available), as a function of 0.5-4.5 keV X-ray flux.

\subsection{Stellar versus extragalactic content}

Despite the high-galactic latitude selection of the {\it XMM-Newton}
fields used in the XMS, a few of our X-ray sources have been
identified as stars.  A detailed study of the stellar content of the
XMS is beyond the scope of this paper, but similarly to what is
found in the {\it XMM-Newton} Galactic Plane Survey (Motch et al.,
in preparation) most of them will be active coronal stars.

The current landscape of X-ray surveys indicates that the stellar
content at high galactic latitudes decreases at faint fluxes.  Since
it is unlikely that any stellar X-ray source has escaped
identification in the XMS survey, we are in a position to quantify
this statement as well as to compare the stellar populations when
selected at different energy bands.

The XMS-X sample contains a total of 20 stars, which represent
$7_{-2}^{+3}\%$ of the sample (henceforth errors on fractions are of
90\% confidence and assuming a binomial distribution). If we split
the XMS-X sample between bright (0.5-4.5 keV flux above $3.3\times
10^{-14}\, {\rm erg}\, {\rm cm}^{-2}\, {\rm s}^{-1}$) and faint
(below the same flux) X-ray sources, the whole sample splits in two
approximately equal halves (143 bright and 141 faint X-ray sources).
The fraction of stars (8) in the faint sample is then
$5.5_{-2.5}^{+4.5}\%$ and the fraction of stars (12) in the bright
sample is $8.5_{-3}^{+4.5}$.

\citet{LopezSantiago07} have explored the stellar content of the BSS
\citep{Dellaceca04}, finding 58/389 ($15\pm 3\%$) stars in the
0.5-4.5 keV sample. Combined with our own measurements on the XMS-X,
this shows that there is a decrease in the stellar content when
going to fainter X-ray fluxes.

It is also interesting to compare the fraction of stars in the
various XMS samples. The soft XMS-S sample contains 15 stars, which
represent $7_{-2}^{+4}\%$ of the sample, similar to the XMS-X.  The
stellar content in the XMS-X and XMS-S samples is very similar.

Stars are much rarer in the XMS-H and XMS-U samples: the XMS-H
contains only 3 stars ($2_{-1}^{+2.5}\%$) and the XMS-U contains no
stars whatsoever ($<4\%$ at 90\% confidence).  In this case we are
totally confident that we are not missing any stars, as all
unidentified sources in the XMS-H and XMS-U samples are optically
extended. The small stellar content in these samples is not a
surprise, as most of our stars are seen in X-rays because of their
active coronae, which have X-ray spectra that are dominated by soft
X-ray line emission and peak around 1 keV.

\subsection{Luminosity and Redshift distributions}

\begin{figure*}
   \centering {\includegraphics[height=8cm,angle=270]{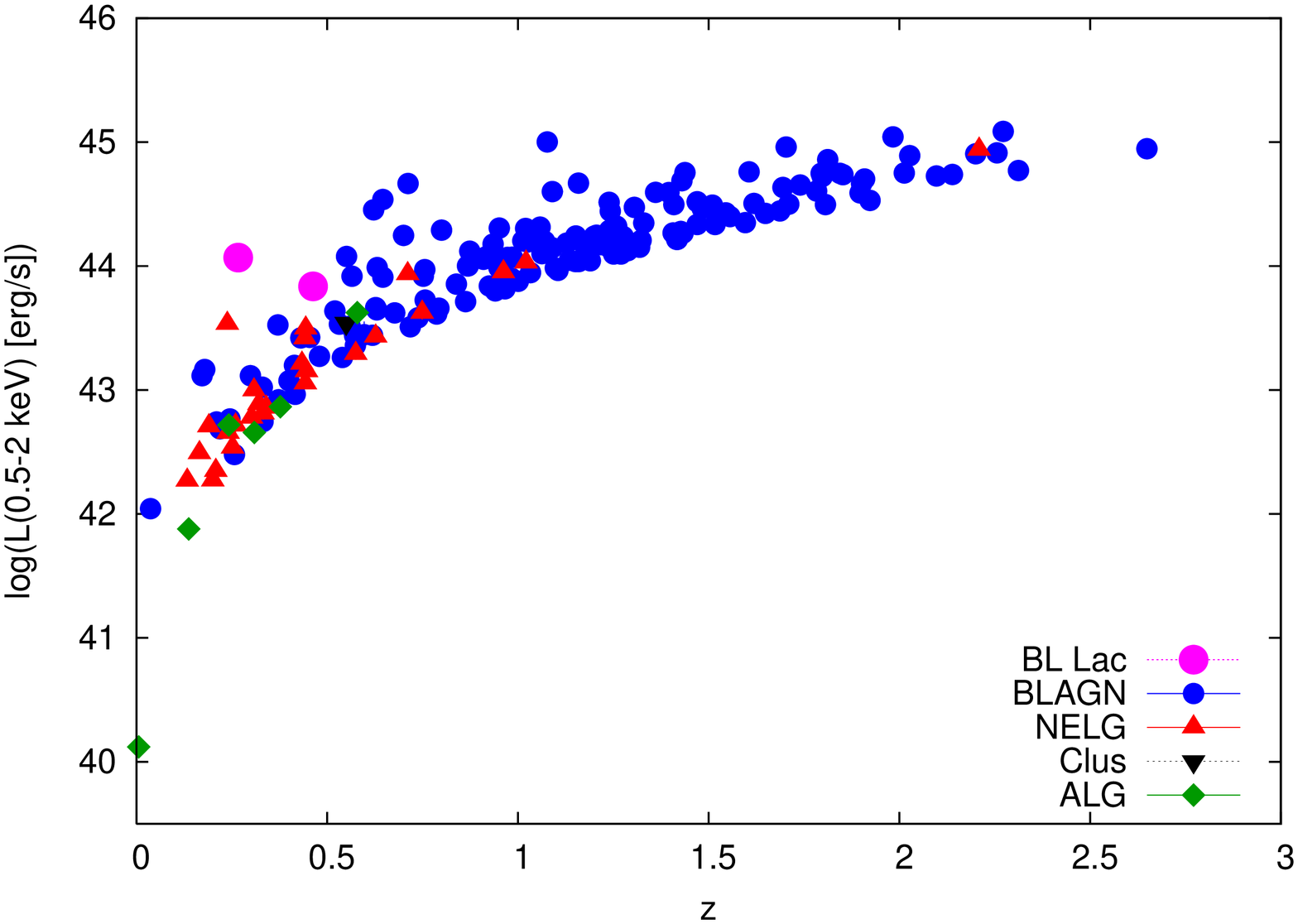}
   \includegraphics[height=8cm,angle=270]{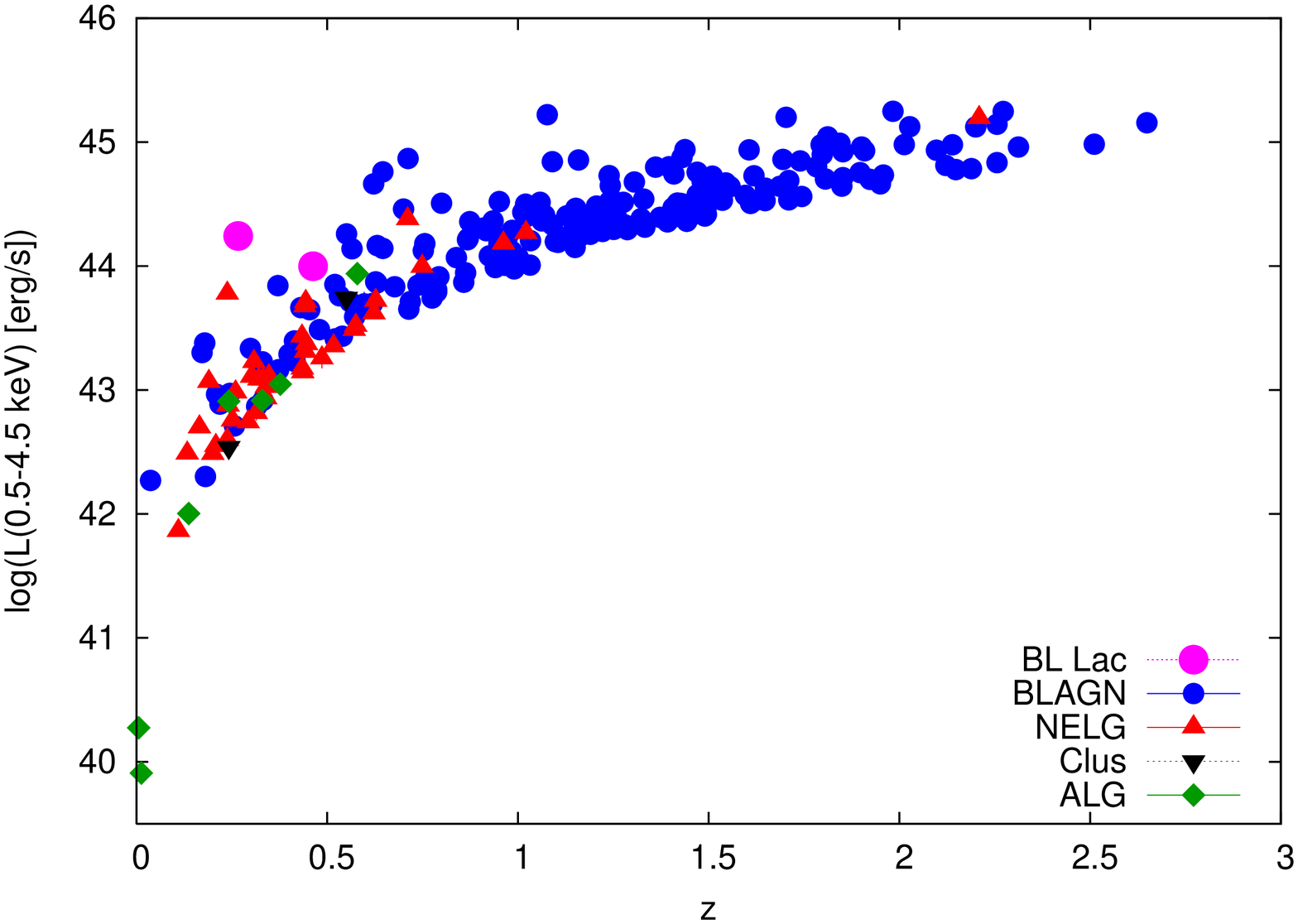}}
   \centering {\includegraphics[height=8cm,angle=270]{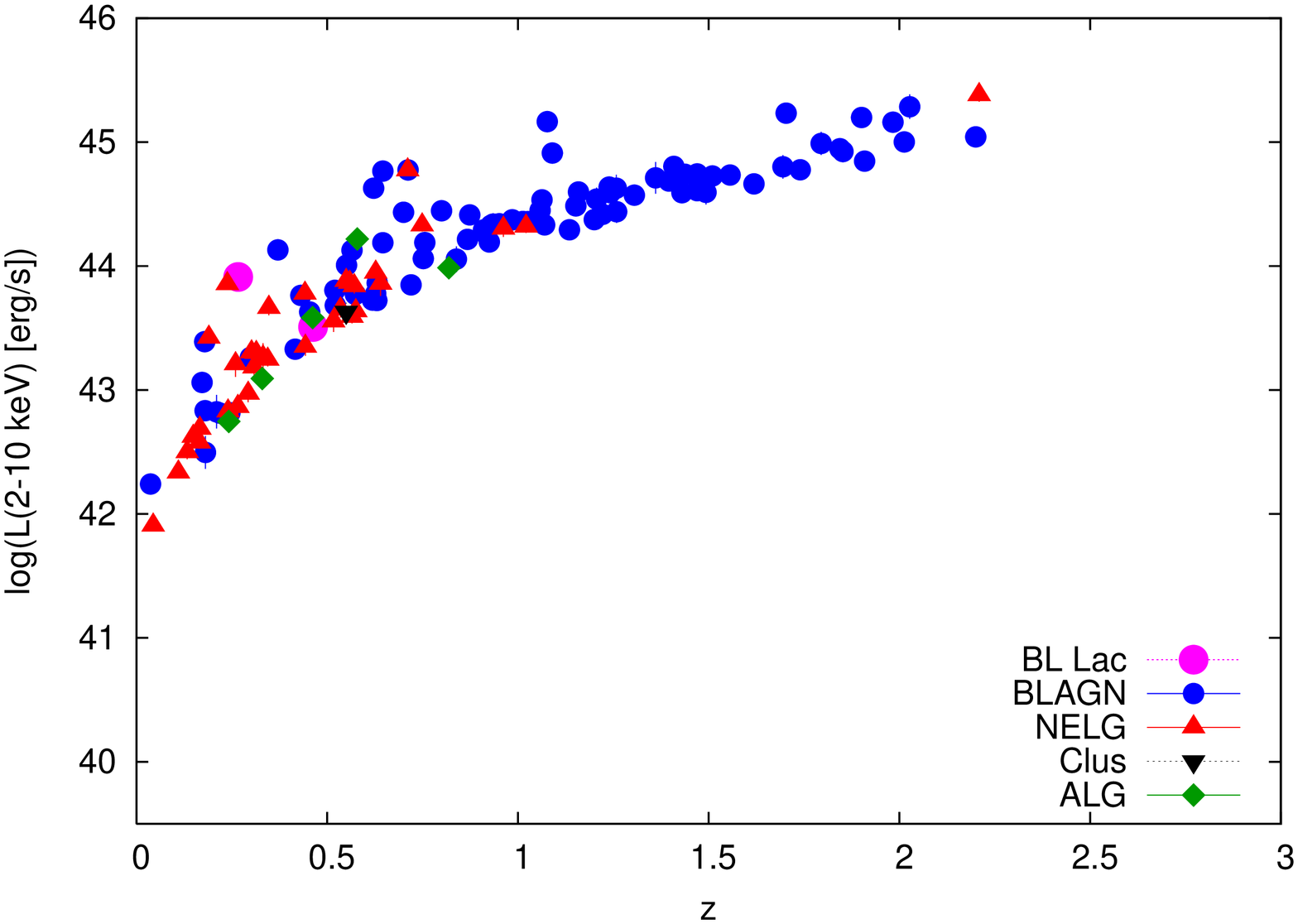}
   \includegraphics[height=8cm,angle=270]{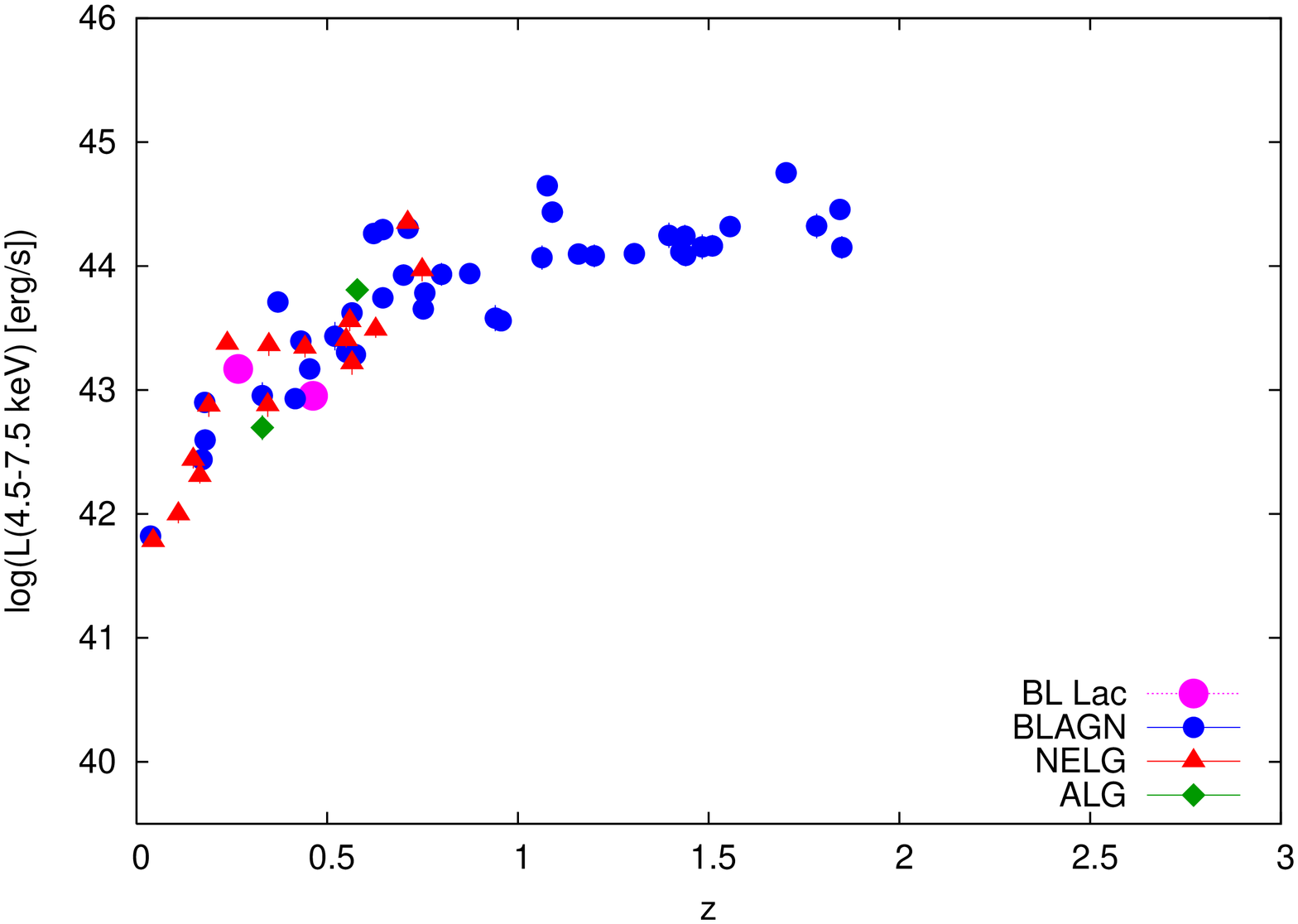}}
      \caption{X-ray luminosity in the selection band versus redshift for extragalactic sources in each one of the XMS samples. Top
      left is for the XMS-S, top right for the XMS-X, bottom left for the XMS-H and
      bottom right for the XMS-U.}
         \label{LXvsz}
   \end{figure*}

The vast majority of XMS sources are extragalactic.  We have
computed the X-ray luminosities of the extragalactic sources (not
corrected for absorption) and these are represented in
Fig.~\ref{LXvsz} as a function of redshift for each of the XMS
samples.

A visual inspection of these $L-z$ relations reveals that all but a
few sources optically classified as NELGs have X-ray luminosities in
the corresponding band in excess of $10^{42}\, {\rm erg}\, {\rm
s}^{-1}$, and are therefore most likely to host a hidden AGN.  With
very few exceptions, NELGs in our survey are therefore type 2 AGN.
In fact, the 2-10 keV luminosity of 6 such objects exceeds
$10^{44}\, {\rm erg}\, {\rm s}^{-1}$ and therefore qualify as type 2
QSOs in standard X-ray astronomy parlance.

The X-ray luminosity of a fraction of sources that we classified as
ALG, also exceeds $10^{42}\, {\rm erg}\, {\rm s}^{-1}$. Specifically
the number of ALG that exceed this luminosity threshold are 4 out of
7 in the most numerous and complete XMS-X. Such sources are often
referred to as X-ray Bright Optically Normal Galaxies (XBONGS) and
when studied in detail are invariably seen to host an AGN, which is
either heavily obscured, or of low luminosity and in any case
outshone by the host galaxy, as shown by \citet{Severgnini03},
\citet{Rigby06} and \citet{Caccianiga07}.

We also find a couple of X-ray sources classified as ALG with low
X-ray luminosities ($\sim 10^{40}\, {\rm erg}\, {\rm s}^{-1}$). In
at least one case (XMSJ 084221.6+705758) the position of the X-ray
source falls in the outskirts of the galaxy, and therefore is likely
to be a candidate for an Ultra-luminous X-ray source.
\citet{Watson05} have discussed some of these sources in the context
of the Subaru/{\it XMM-Newton} Deep Survey.

The overall luminosity distribution in all four samples is centered
around $10^{44}\, {\rm erg}\, {\rm s}^{-1}$, which means that the
sample contains both Seyfert-like AGN and QSOs. This value is also
where the AGN X-ray luminosity function exhibits a knee and
therefore where most of the X-ray volume emissivity comes from.

\begin{figure*}
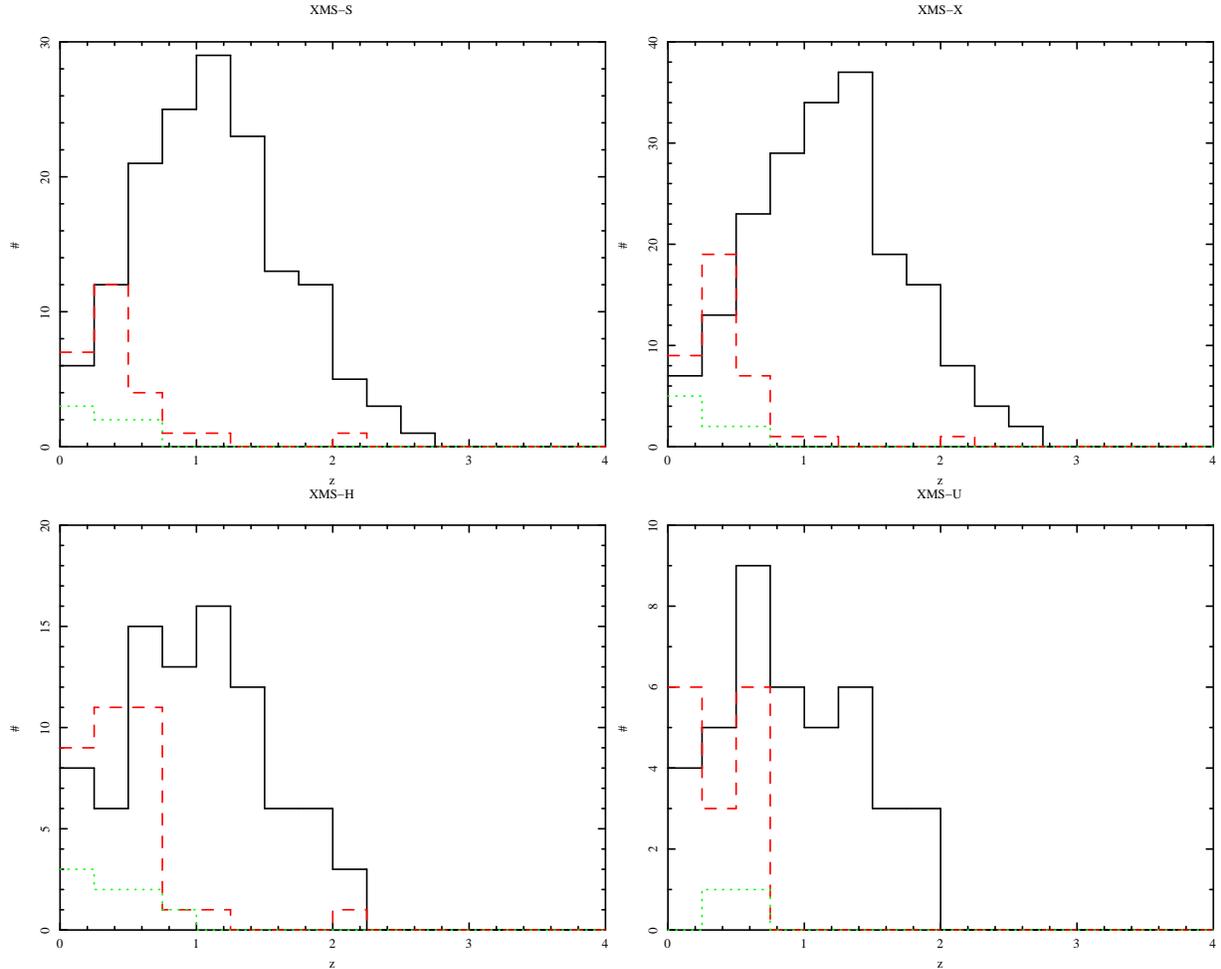

   \centering {\includegraphics[height=8cm,angle=270]{fig5a.ps}
   \includegraphics[height=8cm,angle=270]{fig5b.ps}}
   \centering {\includegraphics[height=8cm,angle=270]{fig5c.ps}
   \includegraphics[height=8cm,angle=270]{fig5d.ps}}
      \caption{Redshift histograms in each one of the XMS samples. Solid line is for BLAGN, dashed
      line for NELG and dotted line for ALG. Top
      left is for the XMS-S, top right for the XMS-X, bottom left for the XMS-H and
      bottom right for the XMS-U.}
         \label{zdist}
   \end{figure*}

The redshift distribution is displayed in Fig.~\ref{zdist} for the
four samples. The peak of the BLAGN population in the XMS-S and
XMS-X samples is around $z\sim 1.5$ which is not far from the one
found in deeper surveys.  However, the redshift cutoff at around
$z\sim 3$ is due to the limited depth of the XMS that fails to find
the higher redshift AGN revealed by deeper surveys.

The contribution from NELG and ALG, most of which are obscured AGN,
peaks at low redshift, typically $z<0.5$.  This is lower than the
peak revealed by deep surveys, due to the modest depth of the
survey. What is worth noting, however, is the comparison between the
various XMS samples. Comparing the redshift distribution for the
softer XMS-S sample to the hard XMS-H sample (which are drawn from a
different parent population according to the Kolmogorov-Smirnov test
which gives a probability of $10^{-13}$ for the null hypothesis)
shows that with a similar sky density the hard sample misses an
important fraction of unobscured AGN (BLAGN) at high redshift but
includes virtually all the obscured objects. The redshift
distribution is consequently shifted to lower values. We next
discuss in more detail the relative fraction of obscured AGN.

\subsection{Obscured versus unobscured AGN}

The fraction of obscured AGN is known to have a strong dependence on
the X-ray selection band and also on the depth of the survey.
Typically soft X-ray selection misses a large fraction of obscured
(and therefore likely absorbed in the X-ray band) AGN.  Deeper X-ray
surveys, even with soft X-ray sensitivity only, have also produced
increasingly large fractions of obscured AGN.

The broad bandpass of {\it XMM-Newton} allows us to study the
fraction of obscured AGN as a function of selection band and depth,
at the intermediate fluxes sampled by the XMS.  A detailed
multi-wavelength study of the XMS survey, combining X-ray spectral
information, optical colours and data at infrared and radio
wavelengths is in preparation (Bussons-Gordo, in preparation).

For the purposes of the current discussion, we now classify as an
AGN any extragalactic X-ray source whose 2-10 keV X-ray luminosity
exceeds $10^{42}\, {\rm erg}\, {\rm s}^{-1}$ and is not obviously
associated with a cluster of galaxies. This stems from the
observational fact that in the local Universe all sources more
luminous than this are at the very least suspected to harbour an
AGN. A potential limitation of our classification, and therefore of
our estimates of the fraction of optically obscured sources among
AGN, comes from the limited quality of the optical spectra, in
particular if weak broad emission lines are present. This is
illustrated in the BSS (Caccianiga et al 2007,Della Ceca et al, in
preparation) where some of the sources originally classified as ALG
or NELG turned out to have ``elusive'' broad emission lines. For the
purposes of this paper, an AGN which does not display an obvious
dominant broad emission line, is considered as an obscured AGN.

The sources that we have classified as BLAGN are unobscured AGN. In
fact, a fraction of these (around 10 per cent) display X-ray
photoelectric absorption \citep{Mateos05a,Mateos05b}, but whatever
the nature of these absorbers, they do not contain enough dust to
obscure the broad line regions of these AGN, and in this context we
will not consider these to be obscured AGN.

Among the sources classified optically as NELG or ALG, an important
fraction of them are AGN according to the above scheme, and we term
these as obscured AGN, since their Broad Line Region is not seen. It
must be stressed anyhow that obscured AGN are expected to follow the
AGN unified model predictions in the sense that they host the same
central engine as an unobscured AGN, but that due to the presence of
dust the Broad Line Region is heavily reddened and therefore not
seen. Photoelectric absorption is expected in their X-ray spectrum
(and certainly seen in most cases), but in about half of AGN without
broad emission lines X-ray absorption is undetected
\citep{Mateos05a}. There are a number of hypotheses that can explain
this mismatch, some of them dealing with the structure of the AGN
itself, and not with real obscuration of a standard AGN
\citep{Mateos05b}. However, this discussion is beyond the scope of
this paper, and we stick to the standard interpretation that the
lack of broad emission lines is equivalent to obscuration.

A potential problem in the study of the fraction of the obscured
objects among the AGN population arises because the fraction of
unidentified X-ray sources is higher for optically fainter sources
and these are more likely to be obscured. There is an indication of
this being true, as most of their optical counterparts appear
extended and therefore dominated by host galaxy light rather than by
the nucleus.

For the XMS-X, we find 42 optically obscured AGN out of a total
sample of 236 identified AGN, which represent $18_{-4}^{+4}\%$. This
fraction is rather robust as it remains virtually unchanged if we
restrict its estimate to the "South" XMS-X sample complete sample
($20_{-5}^{+5}\%$).

The fact that this fraction is indeed much smaller than what is
expected from local Universe studies, where obscured AGN outnumber
unobscured ones by a factor of 3, is indeed due to the fact that
obscuration comes along with photoelectric X-ray absorption which
suppresses X-rays, particularly in the soft band.  This implies that
at harder X-ray energies there should be a higher fraction of
obscured AGN.  Although this is what qualitatively emerges from
existing X-ray surveys, the size and combination of various
selection bands on the XMS can provide a quantitative measurement of
these effects. There is also some qualitative perception that the
fraction of obscured AGN increases substantially when going deeper
in a given X-ray energy band.  In what follows we attempt to test
these statements.

If we divide again the XMS-X sample in two approximately equal
halves of faint and bright X-ray sources (0.5-4.5 keV fluxes below
and above $3.3\times 10^{-14}\, {\rm erg}\, {\rm cm}^{-2}\, {\rm
s}^{-1}$), among the identified sources obscured AGN represent
$17_{-5}^{+6}\%$ of the faint AGN and $19_{-5}^{+6}\%$ of the bright
AGN. This lack of flux dependence, is confirmed when we restrict it
to the "South" XMS-X complete sample ($20_{-6}^{+9}\%$ and
$19_{-6}^{+9}\%$ of obscured AGN for faint and bright sources
respectively). This is within errors of what comes out if we assume
that {\it all} unidentified sources in the XMS-X sample are obscured
AGN, in which case the fraction of obscured AGN would be slightly
higher ($25_{-4}^{+5}\%$) and independent of flux. Incidentally,
comparison of this fraction to the $\sim 20\%$ of obscured AGN in
the South XMS-X sources, shows that despite being an important
obscured AGN component among the unidentified XMS-X sources, there
might be some type 1 AGN among them. We will return to this point
later when discussing X-ray-to-optical flux ratios.

In the XMS-S the fraction of obscured AGN is $17_{-4}^{+5}\%$, which
is very marginally smaller than in the XMS-X sample.  This fraction
remains unchanged when we split the XMS-S in faint and bright
sources.

Things change significantly when we deal with hard X-ray selected
sources. The identified sources in the XMS-H sample contain $35\pm
7\%$ obscured AGN which could be as high as $45\pm 7\%$ if all
unidentified sources are obscured AGN. None of these figures change
between XMS-H bright and faint X-ray sources.

There are no dramatic changes when we use the XMS-U sample with
respect to the XMS-H sample: obscured AGN represent $31_{-6}^{+7}\%$
of the AGN population which might be slightly higher if all
unidentified sources are obscured AGN ($43_{-6}^{+7}\%$).

In summary, in soft X-ray selected samples at intermediate fluxes,
about $\sim 20-25\%$ of the AGN are obscured, and this applies to
both 0.5-2 keV and 0.5-4.5 keV selection. The {\it XMM-Newton} BSS
\citep{Dellaceca04}, which is also selected in the 0.5-4.5 keV band
but at brighter fluxes, finds a slightly smaller fraction of
obscured AGN, in the range of 6 to 14\%.  In the opposite flux
direction, the $ROSAT$ ultra-deep survey \citep{Lehmann01}, which
contains 94 X-ray sources with a 0.5-2 keV flux down to $1.2\times
10^{-15}\, {\rm erg}\, {\rm cm}^{-2}\, {\rm s}^{-1}$ and identified
to 90\% completeness, also found $\sim 20\%$ of obscured objects
among the AGN population (13 out of 70).

The fraction of obscured AGN goes up to $\sim 35-45\%$ for hard
X-ray selected samples at intermediate fluxes in the XMS. This
applies equally to 2-10 keV selection and to 4.5-7.5 keV selection.
This means that above 4.5 keV the sensitivity of {\it XMM-Newton} is
not high enough, and our exposure times are not deep enough, to
raise new heavily obscured X-ray sources that are not selected in
the 2-10 keV band. These fractions do not appear to change with
X-ray flux of the sources within the flux ranges sampled by our
survey. In this case, comparison with the Hard Bright Source Survey
(Caccianiga et al., 2004, Caccianiga et al., 2007, Della Ceca et
al., in preparation), selected in the 4.5-7.5 keV band shows no
change in the fraction of obscured AGN, which these authors quantify
as $31-33\%$.  The {\it Chandra} Multi-wavelength Survey
\citep{Silverman05} (which goes down to 2-10 keV fluxes beween
$10^{-15}$ and $10^{-14}\, {\rm erg}\, {\rm cm}^{-2}\, {\rm
s}^{-1}$), when restricted to optically bright sources, reports that
28\% of the total source sample is obscured, but its identified
fraction in only 77\% and therefore this fraction is most likely a
lower limit.

A final point to address is the dependence of the fraction of
optically obscured AGN as a function of X-ray luminosity. This
fraction is reported by \citet{Barger05} and \citet{Gilli07} among
others, to decrease towards high luminosities.  From
Fig.~\ref{LXvsz} we can see this effect clearly happening in the
XMS. For the XMS-H the fraction of optically obscured extragalactic
objects with 2-10 keV luminosity between $10^{42}$ and $10^{44}\,
{\rm erg}\, {\rm s}^{-1}$ is 62\% (34 out of 55 objects) and above
$10^{44}\, {\rm erg}\, {\rm s}^{-1}$ is only 9\% (6 out of 70
objects). These numbers are consistent, within errors, with those
quoted by \citet{Barger05} and \citet{Gilli07}. However,
Fig.~\ref{LXvsz} itself shows the very restricted coverage of the
luminosity-redshift plane of a single flux-limited survey. We
believe that addressing this issue needs the combination of multiple
surveys covering different depths and solid angles in a way that
samples evenly the luminosity-redshift plane.

\subsection{X-ray to optical flux ratio}

The X-ray to optical flux ratio has been used in various surveys as
a proxy for obscuration. Similarly to other papers
\citep{Krumpe07,Cocchia07}, we use as a proxy for optical flux that
in the $r'$ band and therefore compute $\log
f_{opt}=-0.4r'+\log(f_{r'0}\delta\lambda)$, where
$f_{r'0}=2.40\times 10^{-9}\ {\rm erg}\, {\rm cm}^{-2}\, {\rm
s}^{-1}\, {\rm \AA}^{-1}$ is the zero-point for $r'$ and
$\delta\lambda= 1358\, {\rm \AA}$ is the FWHM of the $r'$ filter.
Note that this yields $X/O=\log(f_X/f_{opt})=\log(f_X)+0.4r'+5.49$,
where $f_X$ is the 2-10 keV flux in $\rm erg\, cm^{-2}\, s^{-1}$,
not corrected for Galactic absorption (the correction is
insignificant at the XMS Galactic latitudes). Typically, unobscured
type 1 AGN have $-1<X/O<1$, and therefore sources with X-ray to
optical flux ratio in excess of 10 have been considered as likely
obscured AGN.

We have excluded from this analysis those sources for which we have
no reliable $r'$ magnitudes, to avoid uncertainties.  We attempted
to calibrate the $R$ versus $r'$ relation, where the $R$ magnitudes
are mostly extracted from the literature and the USNO A2 catalogue.
Specifically, Table~\ref{IDtable} contains 105 BLAGN and 30 NELG for
which we have both $R$ and $r'$. Formally, $\log f_R$ versus $\log
f_{r'}$ yields an offset of $-0.30$ for BLAGN and $-0.49$ for NELG,
but in both cases the scatter is very large (0.25 dex).  Therefore,
by adding into this analysis those sources for which only $R$ is
available, we would be expanding considerably their uncertainties
and therefore we decided to ignore these sources.

\begin{figure*}
   \centering {\includegraphics[height=8cm,angle=270]{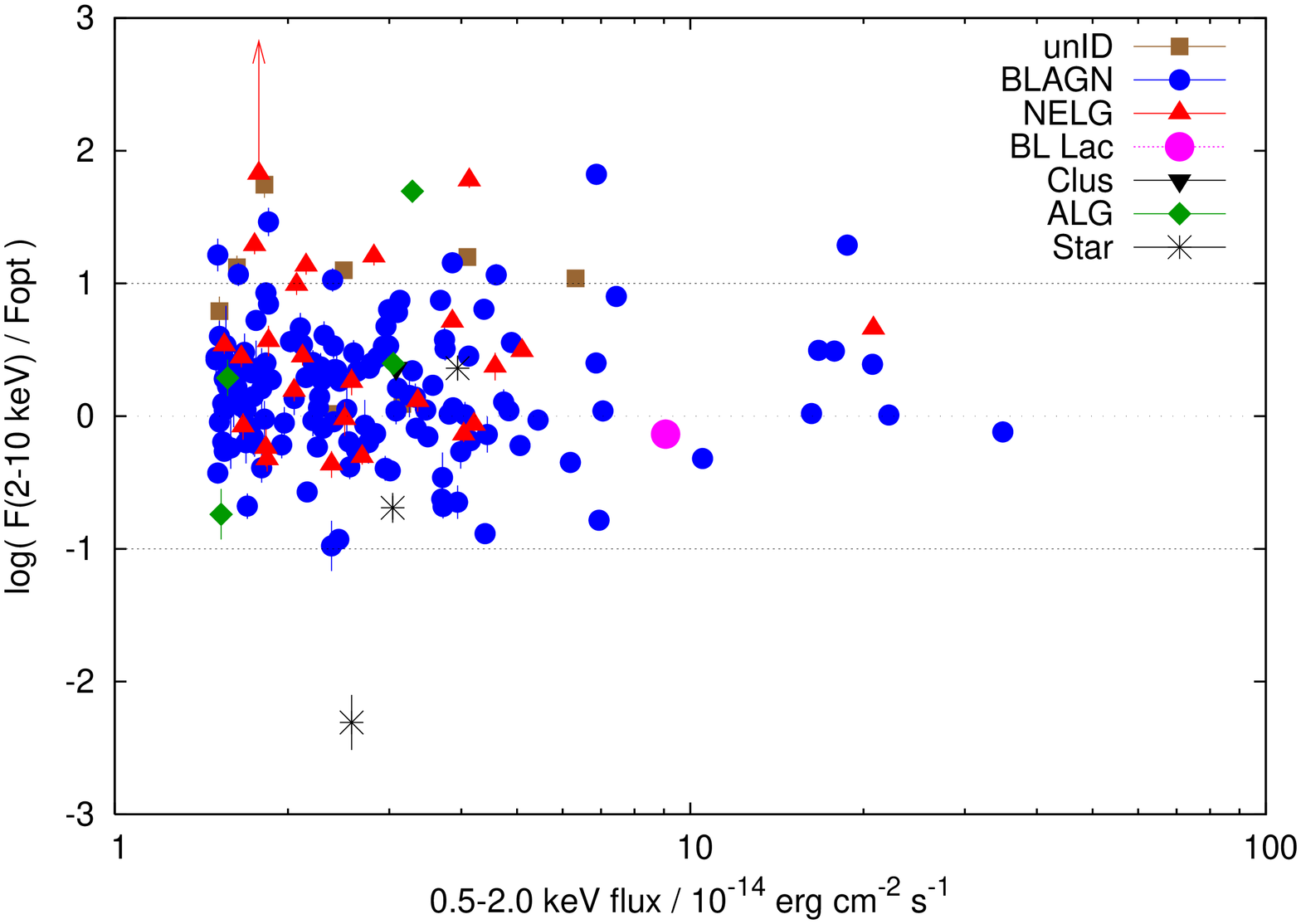}
   \includegraphics[height=8cm,angle=270]{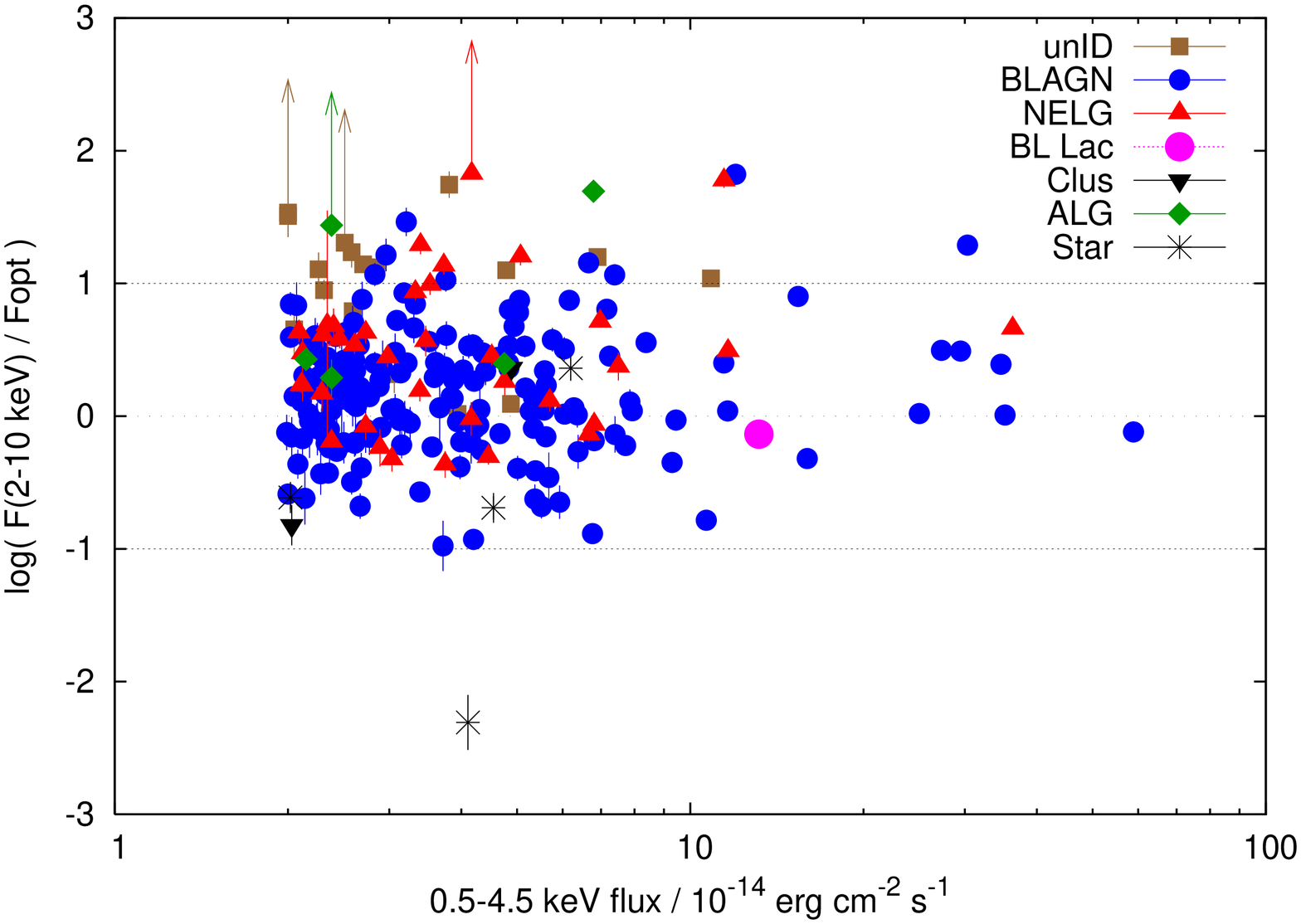}}
   \centering {\includegraphics[height=8cm,angle=270]{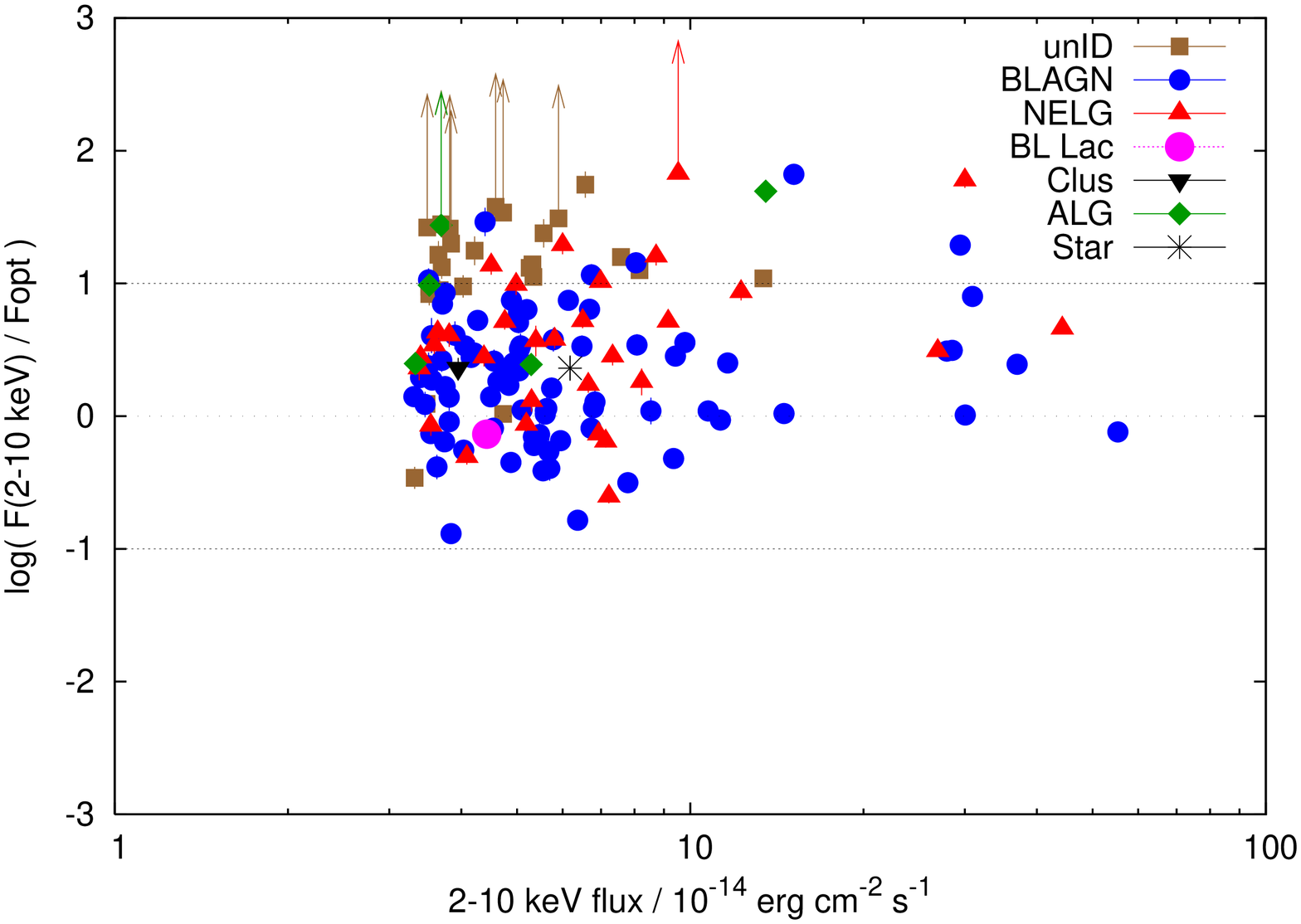}
   \includegraphics[height=8cm,angle=270]{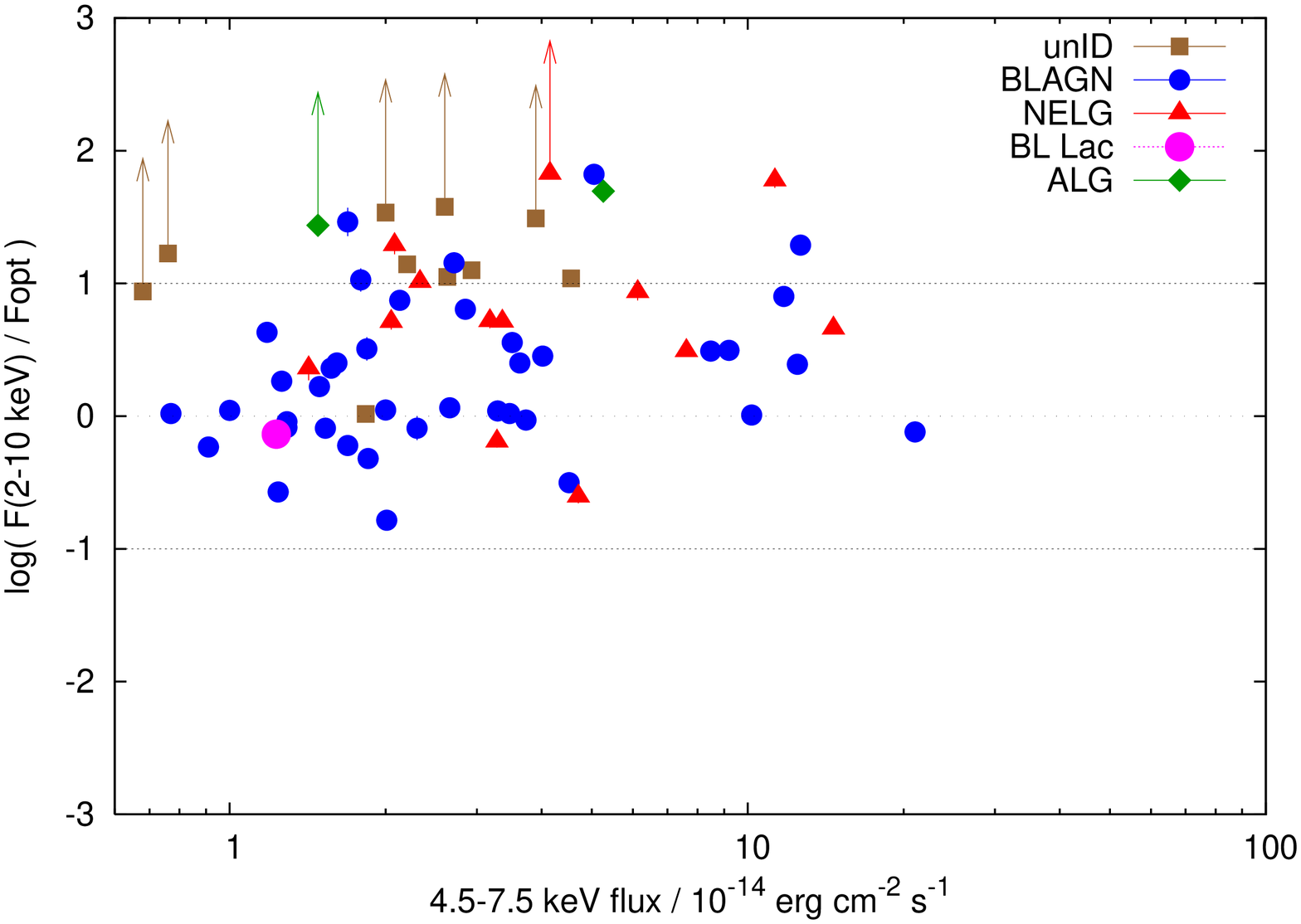}}
      \caption{X-ray to optical flux ratio as a function of X-ray flux in the corresponding
      X-ray band for all 4 XMS samples.  Only those sources with known counterpart, which has
      a measured value (or a lower limit) of $r'$ from our WFC imaging is included.  In this way we miss
      a number of sources in each sample, but we avoid uncertain conversion factors between different bands.
      Top left is for the XMS-S, top right for the XMS-X, bottom left for
the XMS-H and
      bottom right for the XMS-U. Symbols are as in Fig.~\ref{FXFopt_flux_all}}
         \label{FXFopt_FX}
   \end{figure*}

We have studied the fraction of obscured AGN in the various XMS
samples. This is best seen in Fig.~\ref{FXFopt_FX}, where we find
the vast majority of our objects to lie in the ``normal'' type 1 AGN
domain $-1< X/O<1$. However, we also note that a fraction of XMS
sources have extreme values of $ X/O$. Values below $-1$ are usually
dominated by stars.

Far more important is the  other extreme $X/O>1$ where obscured AGN
are expected. Table~\ref{fractions} shows the numbers and fractions
of obscured AGN in the various XMS samples and in 3 ranges of $X/O$.
A difficulty is how we deal with lower limits to optical fluxes of
various sources, where there is no detection in the WFC image, but
only an upper limit in their magnitude from the sensitivity of the
corresponding image. There are a number of uncertainties on this,
including the likely possibility that the undetected counterpart is
optically extended and therefore might be actually brighter (in
integrated magnitude) than the quoted lower limit. For these sources
(which are actually very few) we have used the lower limit in $X/O$
as if it were a real detection.

The first result that becomes evident from Table~\ref{fractions} is
the fact that the fraction of sources with $X/O>1$, and therefore
potentially obscured, varies substantially between samples. For the
XMS-S, we find only 5 sources with $X/O>1$ among a sample of 180
with measured $X/O$ which represent only $3\pm 2\%$. This percentage
grows to $7_{-2}^{+4}\%$ (17/245) for XID-X and grows even further
for the XMS-H to $17_{-5}^{+7}\%$ (23/138) (for the XMS-U the
numbers are too small to reach any conclusion).  It is therefore
clear that the fraction of sources with large $X/O$ goes up from a
small few per cent in the 0.5-2 keV selected XMS-S sample to 20\% in
the 2-10 keV selected XMS-H sample, with XMS-X in between. Note that
these percentages have to be revised slightly downwards, as we are
missing $r'$ magnitudes of most of the stars (more abundant in the
softer samples) which are totally saturated in our WFC images.

\begin{table}
\caption{\label{fractions} Fraction of obscured AGN in the various
XMS samples. ($f_{obsc}$) is the fraction actually measured and
($f_{obsc}^*$) the fraction that would result under the assumption
that all unidentified sources are obscured AGN.}
\begin{tabular}{l c c}
\hline\hline
$\log (F(2-10 keV)/F_{r'})$ & $f_{obsc}$ & $f_{obsc}^*$\\
\hline
             XMS-S\\ \hline
$-1.0:0.0$ & $17_{-6}^{+8}\%\ (14/84)$ & $18_{-6}^{+8}\%\ (15/85)$\\
$0.0:+1.0$ & $17_{-6}^{+8} \%\ (14/81)$& $25_{-7}^{+8}\%\ (22/89)$\\
$>+1.0$    & $25_{-15}^{+49}\%\  (1/4)$ & $40_{-21}^{+41}\%\ (2/5)$\\
\hline
             XMS-X\\ \hline
$-1.0:0.0$ & $12_{-5}^{+9}\%\ (8/65)$ & $14_{-6}^{+8}\%\ (9/66)$\\
$0.0:+1.0$ & $20_{-5}^{+6}\%\ (29/146)$ & $25_{-5}^{+6}\%\ (39/156)$\\
$>+1.0$    & $36_{-17}^{+29}\%\ (4/11)$ & $59_{-17}^{+19}\%\ (10/17)$\\
\hline

             XMS-H\\ \hline
$-1.0:0.0$ & $21_{-9}^{+16}\%\ (6/28)$ & $24_{-10}^{+16}\%\ (7/29)$\\
$0.0:+1.0$ & $28_{-7}^{+9}$\%\ (22/78) & $34_{-8}^{+9}\%\ (29/85)$\\
$>+1.0$    & $50_{-18}^{+25}\%\ (6/12)$ & $74_{-14}^{+14}\%\ (17/23)$\\
\hline
             XMS-U \\ \hline
$-1.0:0.0$ & $14_{-7}^{+15}\%\ (4/29)$ & $17_{-8}^{+17}\%\ (5/30)$\\
$0.0:+1.0$ & $39_{-15}^{+22}\%\ (7/18)$ & $50_{-14}^{+19}\%\ (11/22)$\\
$>+1.0$    & $50_{-25}^{+40}\%\ (2/4)$ & $50_{-25}^{+40}\%\ (2/4)$\\
\hline\hline
\end{tabular}
\end{table}

We see from Table~\ref{fractions} that the fraction of obscured AGN
among the $X/O>1$ sources is higher than in the whole sample, and
this holds for all XMS samples.  The second fact that can be seen by
inspecting Table~\ref{fractions} is that in the XMS-H the fraction
of obscured AGN amongst the $X/O>1$ sources could be as high as 90\%
if all unidentified sources are obscured AGN, but this percentage is
lower for the XMS-S and XMS-X.

But we can also look at this fact from a different point of view,
which is that there are unobscured AGN with $X/O>1$. There are at
least 3/5, 7/17, 6/23 and 2/4 unobscured AGN with $X/O>1$ in the
XMS-S, XMS-X, XMS-H and XMS-U samples respectively, which represent
somewhere between one fifth and one half of the corresponding sample
with a selection cut at $X/O>1$.  The nature of these
 BLAGN with $X/O>1$ will be investigated in future papers.

\section{Conclusions}
\label{sec:conclusions}

In this paper we have presented the {\it XMM-Newton} Medium
sensitivity Survey XMS, and extracted a number of robust
quantitative conclusions about the population of high Galactic
latitude X-ray sources at intermediate flux levels. We have argued
that given the completeness of our identifications and the
relatively large size of the XMS samples, these conclusions can be
safely exported to a much larger X-ray source catalogue like 2XMM.

Our conclusions can be summarized as follows:

\begin{enumerate}

\item The high galactic latitude X-ray sky at intermediate flux
levels is dominated by AGN, which includes type-1 and type-2 AGN as
well as the so-called XBONG which are likely to host a low
luminosity or obscured nucleus (or both).  The stellar content is
less than 10\% in soft X-ray selected samples, and drops to below
5\% at around soft X-ray fluxes $\sim 10^{-14}\, {\rm erg}\, {\rm
cm}\, {\rm s}^{-1}$. The stellar content in hard X-ray selected
samples does not exceed a few per cent at most. Selection in 0.5-4.5
keV produces intermediate results.

\item Given the limited sensitivity of {\it XMM-Newton} above a few
keV -which is due to the roll over of effective area- current
surveys conducted in the so-called ultra-hard band (4.5-7.5 keV) do
not bring any new source population or any significant difference
with respect to 2-10 keV selected surveys. Much longer exposure
times would be needed to unveil any new heavily obscured population
with {\it XMM-Newton}.

\item Obscured AGN represent $\sim 20\%$ of the soft X-ray selected
population of AGN, all the way from  $\sim 10^{-13}\, {\rm erg}\,
{\rm cm}^{-2}\, {\rm s}^{-1}$ down to  $\sim 10^{-15}\, {\rm erg}\,
{\rm cm}^{-2}\, {\rm s}^{-1}$, with no compelling evidence for an
increase of this fraction towards fainter fluxes within this range.

\item Likewise, obscured AGN represent $\sim 35\%$ ($45\%$
if all unidentified sources are obscured AGN) of the hard X-ray
selected population of AGN, with no hint of an increase down to a
hard X-ray flux  $\sim 10^{-14}\, {\rm erg}\, {\rm cm}^{-2}\, {\rm
s}^{-1}$.

\item The fraction of X-ray sources with X-ray to optical flux ratio
$>10$ (or $X/O>1$ using the notation of this paper) is a mere 3\% in
soft X-ray selected samples, but grows to 20\% in hard X-ray
selected samples.

\item Those sources with $X/O>1$ are mostly obscured AGN, but a
fraction of around  20\% of them in the hard band are unobscured
type-1 AGN. This means that $X/O>1$ alone cannot be used as a proxy
for obscured X-ray sources.

\end{enumerate}

\begin{acknowledgements}
We are grateful to the International Scientific Committee of the
Canary Islands' observatories for a generous allocation of observing
time in 2000 and 2001, through the International Time Programme
scheme. We are grateful to the Calar Alto Time Allocation Committee
for continued support to the optical spectroscopic identification
programme. Authors at the Instituto de F\'\i sica de Cantabria (XB,
FJC, MTC, JB-G, AC, JE and FP) acknowledge financial support by the
Spanish Ministerio de Educaci\'on y Ciencia under projects
ESP2003-00812 and ESP2006-13608-C02-01. We thank J.L. Mui\~nos and
D.W. Evans for help with the CMC survey. AC, RDC, TM and PS
acknowledge financial support from the Italian Space Agency (ASI),
the Ministero dell'Universita´ e della Ricerca (MIUR) and Istituto
Nazionale di Astrofisica (INAF) over the last few years. This work
was supported by the German DLR under contract 50 OR 0201.

\end{acknowledgements}

\voffset=7truecm
\begin{landscape}
\centering
\longtab{2}{
\begin{longtable}{l c c c l c c c c c}
\caption{\label{Xraysources}XMS X-ray source list}\\
 \hline\hline
Source Name &RA (J2000) & DEC (J2000) &$r_{90}^a$     &Sample &$S(0.5-2)^b$   & $S(2-10)^b$   &$S(0.5-4.5)^b$ & $S(4.5-7.5)^b$    & $HR_2^c$\\

\hline
\endfirsthead
\caption{continued.}\\
\hline\hline
Source Name & RA (J2000) & DEC (J2000) &$r_{90}^a$     &Sample &$S(0.5-2)^b$   & $S(2-10)^b$   &$S(0.5-4.5)^b$ & $S(4.5-7.5)^b$    & $HR_2^c$\\

\hline
\endhead
\hline
\endfoot
\hline\hline

\object{XMSJ 001744.6+162815}   & $00:17:44.6$  & $+16:28:15.3$ & $1.0$ & \ H\  \   & $0.18 \pm 0.05$   & $3.54 \pm 0.64$   & $0.86 \pm 0.16$   & $1.17 \pm 0.31$   & $+0.44 \pm 0.14$ \\
\object{XMSJ 001800.9+163626}   & $00:18:00.9$  & $+16:36:26.2$ & $0.7$ & \ HX\     & $1.36 \pm 0.17$   & $3.70 \pm 0.74$   & $2.50 \pm 0.28$   & $1.10 \pm 0.33$   & $-0.45 \pm 0.11$ \\
\object{XMSJ 001802.7+161642}   & $00:18:02.7$  & $+16:16:42.2$ & $0.3$ & S\  X\    & $1.74 \pm 0.18$   & $1.95 \pm 0.59$   & $2.77 \pm 0.27$   & $0.54 \pm 0.30$   & $-0.68 \pm 0.08$ \\
\object{XMSJ 001817.3+161740}   & $00:18:17.3$  & $+16:17:40.1$ & $0.5$ & SHX\      & $3.08 \pm 0.23$   & $3.95 \pm 0.67$   & $4.89 \pm 0.34$   & $1.33 \pm 0.44$   & $-0.68 \pm 0.05$ \\
\object{XMSJ 001821.7+161943}   & $00:18:21.7$  & $+16:19:43.7$ & $0.1$ & S\  X\    & $1.62 \pm 0.14$   & $1.40 \pm 0.29$   & $2.53 \pm 0.20$   & $0.14 \pm 0.06$   & $-0.70 \pm 0.05$ \\
\object{XMSJ 001831.9+162926}   & $00:18:31.9$  & $+16:29:26.0$ & $0.2$ & SHXU      & $10.50 \pm 0.29$  & $9.35 \pm 0.63$   & $15.96 \pm 0.41$  & $1.85 \pm 0.34$   & $-0.74 \pm 0.02$ \\
\object{XMSJ 001837.2+163446}   & $00:18:37.2$  & $+16:34:46.1$ & $0.6$ & \ \  X\   & $1.38 \pm 0.13$   & $1.62 \pm 0.41$   & $2.08 \pm 0.20$   & $0.63 \pm 0.19$   & $-0.76 \pm 0.06$ \\
\object{XMSJ 001838.3+162007}   & $00:18:38.3$  & $+16:20:07.1$ & $0.4$ & S\  X\    & $2.41 \pm 0.15$   & $2.96 \pm 0.41$   & $3.70 \pm 0.22$   & $1.21 \pm 0.32$   & $-0.74 \pm 0.04$ \\
\hline\hline
\end{longtable}
\noindent $^a$ Statistical error radius at 90\% confidence in the
position of the X-ray source in arcsec.

\noindent $^b$ All fluxes are quoted in units of $10^{-14}\, {\rm
erg}\, {\rm cm}^{-2}\, {\rm s}^{-1}$ in the corresponding band

\noindent $^c$ Hardness Ratio defined as $HR_2=(H-S)/(H+S)$, where
$S$ and $H$ are exposure-time corrected count rates in the 0.5-2 keV
and 2-4.5 keV bands respectively.
\end{landscape}
}

\longtab{5}{
\begin{landscape}
\begin{longtable}{l c c c c c c l l c}
\caption{\label{IDtable}Optical identifications of the XMS}\\
\hline\hline
X-ray source        & $RA_o$ (J2000)& $DEC_o$(J2000)& $g'$      & $r'$      & $i'$      & $R$   & Id$^a$   & Type & $z$\\
\hline
\endfirsthead
\caption{continued.}\\
\hline\hline
X-ray source        & $RA_o$ (J2000)& $DEC_o$(J2000)& $g'$      & $r'$      & $i'$      & $R$   & Id$^a$   & Type & $z$\\
\hline
\endhead
\hline
\endfoot
\hline\hline
\object{XMSJ 001744.6+162815}   & $00:17:44.6$  & $+16:28:18.1$ & $20.98$   & $19.73$   & $19.16$   & $18.76$   & Y & NELG  & $0.266$\\
\object{XMSJ 001800.9+163626}   & $00:18:00.9$  & $+16:36:25.4$ & $21.57$   & $20.91$   & $20.65$   &       & Y & BLAGN & $1.222$\\
\object{XMSJ 001802.7+161642}   & $00:18:02.7$  & $+16:16:42.7$ & $20.58$   & $20.16$   & $20.02$   & $19.30$   & Y & BLAGN & $1.407$\\
\object{XMSJ 001817.3+161740}   & $00:18:17.0$  & $+16:17:40.0$ & $22.61$   & $20.70$   & $19.22$   & $19.85$   & Y & Clus  & $0.550$\\
\object{XMSJ 001821.7+161943}   & $00:18:21.8$  & $+16:19:43.6$ & $22.14$   & $21.61$   & $21.26$   &       & Y & BLAGN & $1.160$\\
\object{XMSJ 001831.9+162926}   & $00:18:32.0$  & $+16:29:26.5$ & $18.28$   & $18.06$   & $17.80$   & $18.40$   & Y & BLAGN & $0.551$\\
\object{XMSJ 001837.2+163446}   & $00:18:37.3$  & $+16:34:46.8$ & $20.25$   & $19.86$   & $19.48$   &       & Y & BLAGN & $2.148$\\
\object{XMSJ 001838.3+162007}   & $00:18:38.2$  & $+16:20:07.5$ & $21.83$   & $20.98$   & $21.06$   & $19.40$   & Y & BLAGN & $1.090$\\
\hline\hline
\end{longtable}
\noindent $^a$ This flag means: Y if the source has been positively
identified via optical spectroscopy; C if there is a single clear
optical candidate counterpart from in optical images; E if there is
no candidate counterpart down to our imaging sensitivity. }
\end{landscape}

\end{document}